\documentclass[conference]{IEEEtran}

\usepackage{lipsum}

\usepackage{graphicx}
\usepackage{verbatim}
\usepackage{booktabs}
\usepackage{array}
\usepackage{hyperref}
\usepackage{numprint}
\usepackage{wrapfig}
\usepackage{framed}
\usepackage{rotating}
\usepackage[]{examplep}
\usepackage{stfloats}
\usepackage{amssymb}
\usepackage{amsmath}
\usepackage{enumerate}
\usepackage{threeparttable}

\npthousandsep{,}
\npdecimalsign{.}

\newenvironment{punchline}[1]{
\begin{oframed}
#1
\end{oframed}
}

\newcolumntype{v}[1]{>{\raggedright\hspace{0pt}}p{#1}}

\newcommand{\freqCorp}{165}

\newcommand{\freqComp}{56}

\newcommand{\freqQues}{24}

\newcommand{\freqExpe}{19}

\newcommand{\freqLitSurv}{5}

\newcommand{\existsEqualOneNumLang}{147}

\newcommand{\existsEqualTwoNumLang}{19}

\newcommand{\medianNumCommits}{3292}

\newcommand{\medianNumComparedWithNoCurrent}{3}

\newcommand{\minNumExpParticipants}{2}
\newcommand{\firstQuartileNumExpParticipants}{5}
\newcommand{\medianNumExpParticipants}{16}

\newcommand{\thirdQuartileNumExpParticipants}{34}
\newcommand{\maxNumExpParticipants}{128}

\newcommand{\medianNumExpTasks}{4}

\newcommand{\medianNumLSPapers}{35}

\newcommand{\medianNumLSPapersStart}{2161}

\newcommand{\existsEqualOneNumProjects}{45}
\newcommand{\valueMoreThanTenNumProjects}{10}
\newcommand{\existsMoreThanTenNumProjects}{24}

\newcommand{\existsLessOrEqThreeNumProjects}{99}

\newcommand{\medianNumQsAll}{20}

\newcommand{\medianNumQParticipants}{12}

\newcommand{\medianNumReleases}{10}

\newcommand{\medianNumRevisions}{18870}

\newcommand{\medianNumToolsInPaper}{2}

\newcommand{\maxNumToolsInPaper}{6}

\newcommand{\medianNumVersions}{10}

\newcommand{\medianNumYears}{8}

\newcommand{\minRatioQAllParticipants}{5}

\newcommand{\minRatioLSPapers}{0}

\newcommand{\maxRatioLSPapers}{39}

\newcommand{\medianRatioQParticipants}{19}

\newcommand{\pMaxActInCSMR}{20}

\newcommand{\pMaxActInESEM}{8}

\newcommand{\pMaxActInICPC}{14}

\newcommand{\pMaxActInICSM}{19}

\newcommand{\pMaxActInMSR}{11}

\newcommand{\pMaxActInSCAM}{11}

\newcommand{\pMaxActInWCRE}{19}

\newcommand{\pPapersWithRepPackageOne}{26}
\newcommand{\cPapersWithRepPackageOne}{SCAM}
\newcommand{\pPapersWithRepPackageTwo}{31}
\newcommand{\cPapersWithRepPackageTwo}{ICSM}
\newcommand{\pPapersWithRepPackageThree}{33}
\newcommand{\cPapersWithRepPackageThree}{CSMR}
\newcommand{\pPapersWithRepPackageFour}{33}
\newcommand{\cPapersWithRepPackageFour}{MSR}
\newcommand{\pPapersWithRepPackageFive}{33}
\newcommand{\cPapersWithRepPackageFive}{WCRE}
\newcommand{\pPapersWithRepPackageSix}{38}
\newcommand{\cPapersWithRepPackageSix}{ESEM}
\newcommand{\pPapersWithRepPackageSeven}{48}
\newcommand{\cPapersWithRepPackageSeven}{ICPC}

\newcommand{\pPapersCanReproduceOne}{3}
\newcommand{\cPapersCanReproduceOne}{ICSM}
\newcommand{\pPapersCanReproduceTwo}{4}
\newcommand{\cPapersCanReproduceTwo}{WCRE}
\newcommand{\pPapersCanReproduceThree}{16}
\newcommand{\cPapersCanReproduceThree}{SCAM}
\newcommand{\pPapersCanReproduceFour}{19}
\newcommand{\cPapersCanReproduceFour}{ICPC}
\newcommand{\pPapersCanReproduceFive}{25}
\newcommand{\cPapersCanReproduceFive}{ESEM}
\newcommand{\pPapersCanReproduceSix}{27}
\newcommand{\cPapersCanReproduceSix}{CSMR}
\newcommand{\pPapersCanReproduceSeven}{33}
\newcommand{\cPapersCanReproduceSeven}{MSR}

\newcommand{\pPapersWithComparisonOne}{19}
\newcommand{\cPapersWithComparisonOne}{ICPC}
\newcommand{\pPapersWithComparisonTwo}{22}
\newcommand{\cPapersWithComparisonTwo}{MSR}
\newcommand{\pPapersWithComparisonThree}{26}
\newcommand{\cPapersWithComparisonThree}{SCAM}
\newcommand{\pPapersWithComparisonFour}{30}
\newcommand{\cPapersWithComparisonFour}{WCRE}
\newcommand{\pPapersWithComparisonFive}{33}
\newcommand{\cPapersWithComparisonFive}{ESEM}
\newcommand{\pPapersWithComparisonSix}{36}
\newcommand{\cPapersWithComparisonSix}{ICSM}
\newcommand{\pPapersWithComparisonSeven}{47}
\newcommand{\cPapersWithComparisonSeven}{CSMR}

\newcommand{\pPapersWithHighConfidenceLevelOne}{15}
\newcommand{\cPapersWithHighConfidenceLevelOne}{WCRE}
\newcommand{\pPapersWithHighConfidenceLevelTwo}{26}
\newcommand{\cPapersWithHighConfidenceLevelTwo}{SCAM}
\newcommand{\pPapersWithHighConfidenceLevelThree}{42}
\newcommand{\cPapersWithHighConfidenceLevelThree}{ICSM}
\newcommand{\pPapersWithHighConfidenceLevelFour}{52}
\newcommand{\cPapersWithHighConfidenceLevelFour}{ICPC}
\newcommand{\pPapersWithHighConfidenceLevelFive}{54}
\newcommand{\cPapersWithHighConfidenceLevelFive}{ESEM}
\newcommand{\pPapersWithHighConfidenceLevelSix}{60}
\newcommand{\cPapersWithHighConfidenceLevelSix}{CSMR}
\newcommand{\pPapersWithHighConfidenceLevelSeven}{78}
\newcommand{\cPapersWithHighConfidenceLevelSeven}{MSR}

\newcommand{\pPapersWithLowConfidenceLevelOne}{0}
\newcommand{\cPapersWithLowConfidenceLevelOne}{ICPC}
\newcommand{\pPapersWithLowConfidenceLevelTwo}{6}
\newcommand{\cPapersWithLowConfidenceLevelTwo}{ICSM}
\newcommand{\pPapersWithLowConfidenceLevelThree}{6}
\newcommand{\cPapersWithLowConfidenceLevelThree}{MSR}
\newcommand{\pPapersWithLowConfidenceLevelFour}{7}
\newcommand{\cPapersWithLowConfidenceLevelFour}{CSMR}
\newcommand{\pPapersWithLowConfidenceLevelFive}{13}
\newcommand{\cPapersWithLowConfidenceLevelFive}{ESEM}
\newcommand{\pPapersWithLowConfidenceLevelSix}{16}
\newcommand{\cPapersWithLowConfidenceLevelSix}{SCAM}
\newcommand{\pPapersWithLowConfidenceLevelSeven}{19}
\newcommand{\cPapersWithLowConfidenceLevelSeven}{WCRE}

\newcommand{\pPapersWithModerateConfidenceLevelOne}{17}
\newcommand{\cPapersWithModerateConfidenceLevelOne}{MSR}
\newcommand{\pPapersWithModerateConfidenceLevelTwo}{33}
\newcommand{\cPapersWithModerateConfidenceLevelTwo}{CSMR}
\newcommand{\pPapersWithModerateConfidenceLevelThree}{33}
\newcommand{\cPapersWithModerateConfidenceLevelThree}{ESEM}
\newcommand{\pPapersWithModerateConfidenceLevelFour}{43}
\newcommand{\cPapersWithModerateConfidenceLevelFour}{ICPC}
\newcommand{\pPapersWithModerateConfidenceLevelFive}{53}
\newcommand{\cPapersWithModerateConfidenceLevelFive}{ICSM}
\newcommand{\pPapersWithModerateConfidenceLevelSix}{58}
\newcommand{\cPapersWithModerateConfidenceLevelSix}{SCAM}
\newcommand{\pPapersWithModerateConfidenceLevelSeven}{67}
\newcommand{\cPapersWithModerateConfidenceLevelSeven}{WCRE}

\newcommand{\numPapersWithSourcesCSMR}{9}
\newcommand{\numPapersWithSourcesESEM}{3}
\newcommand{\numPapersWithSourcesICPC}{5}
\newcommand{\numPapersWithSourcesICSM}{9}
\newcommand{\numPapersWithSourcesMSR}{7}
\newcommand{\numPapersWithSourcesSCAM}{3}
\newcommand{\numPapersWithSourcesWCRE}{7}
\newcommand{\pPapersWithSourcesCSMR}{30}
\newcommand{\pPapersWithSourcesESEM}{13}
\newcommand{\pPapersWithSourcesICPC}{24}
\newcommand{\pPapersWithSourcesICSM}{25}
\newcommand{\pPapersWithSourcesMSR}{39}
\newcommand{\pPapersWithSourcesSCAM}{16}
\newcommand{\pPapersWithSourcesWCRE}{26}
\newcommand{\pPapersWithSourcesOne}{13}
\newcommand{\cPapersWithSourcesOne}{ESEM}
\newcommand{\pPapersWithSourcesTwo}{16}
\newcommand{\cPapersWithSourcesTwo}{SCAM}
\newcommand{\pPapersWithSourcesThree}{24}
\newcommand{\cPapersWithSourcesThree}{ICPC}
\newcommand{\pPapersWithSourcesFour}{25}
\newcommand{\cPapersWithSourcesFour}{ICSM}
\newcommand{\pPapersWithSourcesFive}{26}
\newcommand{\cPapersWithSourcesFive}{WCRE}
\newcommand{\pPapersWithSourcesSix}{30}
\newcommand{\cPapersWithSourcesSix}{CSMR}
\newcommand{\pPapersWithSourcesSeven}{39}
\newcommand{\cPapersWithSourcesSeven}{MSR}

\newcommand{\numPapersWithDatasetsCSMR}{5}
\newcommand{\numPapersWithDatasetsESEM}{0}
\newcommand{\numPapersWithDatasetsICPC}{2}
\newcommand{\numPapersWithDatasetsICSM}{6}
\newcommand{\numPapersWithDatasetsMSR}{5}
\newcommand{\numPapersWithDatasetsSCAM}{0}
\newcommand{\numPapersWithDatasetsWCRE}{2}

\newcommand{\pPapersWithEvolutionOne}{14}
\newcommand{\cPapersWithEvolutionOne}{ICPC}
\newcommand{\pPapersWithEvolutionTwo}{16}
\newcommand{\cPapersWithEvolutionTwo}{SCAM}
\newcommand{\pPapersWithEvolutionThree}{21}
\newcommand{\cPapersWithEvolutionThree}{ESEM}
\newcommand{\pPapersWithEvolutionFour}{33}
\newcommand{\cPapersWithEvolutionFour}{CSMR}
\newcommand{\pPapersWithEvolutionFive}{33}
\newcommand{\cPapersWithEvolutionFive}{ICSM}
\newcommand{\pPapersWithEvolutionSix}{33}
\newcommand{\cPapersWithEvolutionSix}{WCRE}
\newcommand{\pPapersWithEvolutionSeven}{56}
\newcommand{\cPapersWithEvolutionSeven}{MSR}

\newcommand{\pPapersWithExperimentOne}{0}
\newcommand{\cPapersWithExperimentOne}{MSR}
\newcommand{\pPapersWithExperimentTwo}{0}
\newcommand{\cPapersWithExperimentTwo}{SCAM}
\newcommand{\pPapersWithExperimentThree}{3}
\newcommand{\cPapersWithExperimentThree}{CSMR}
\newcommand{\pPapersWithExperimentFour}{7}
\newcommand{\cPapersWithExperimentFour}{WCRE}
\newcommand{\pPapersWithExperimentFive}{8}
\newcommand{\cPapersWithExperimentFive}{ICSM}
\newcommand{\pPapersWithExperimentSix}{21}
\newcommand{\cPapersWithExperimentSix}{ESEM}
\newcommand{\pPapersWithExperimentSeven}{38}
\newcommand{\cPapersWithExperimentSeven}{ICPC}

\newcommand{\pPapersWithLitSurveyOne}{0}
\newcommand{\cPapersWithLitSurveyOne}{ICSM}
\newcommand{\pPapersWithLitSurveyTwo}{0}
\newcommand{\cPapersWithLitSurveyTwo}{MSR}
\newcommand{\pPapersWithLitSurveyThree}{0}
\newcommand{\cPapersWithLitSurveyThree}{SCAM}
\newcommand{\pPapersWithLitSurveyFour}{0}
\newcommand{\cPapersWithLitSurveyFour}{WCRE}
\newcommand{\pPapersWithLitSurveyFive}{3}
\newcommand{\cPapersWithLitSurveyFive}{CSMR}
\newcommand{\pPapersWithLitSurveySix}{5}
\newcommand{\cPapersWithLitSurveySix}{ICPC}
\newcommand{\pPapersWithLitSurveySeven}{13}
\newcommand{\cPapersWithLitSurveySeven}{ESEM}

\newcommand{\pPapersWithManualEffortOne}{13}
\newcommand{\cPapersWithManualEffortOne}{CSMR}
\newcommand{\pPapersWithManualEffortTwo}{13}
\newcommand{\cPapersWithManualEffortTwo}{ESEM}
\newcommand{\pPapersWithManualEffortThree}{22}
\newcommand{\cPapersWithManualEffortThree}{ICSM}
\newcommand{\pPapersWithManualEffortFour}{26}
\newcommand{\cPapersWithManualEffortFour}{SCAM}
\newcommand{\pPapersWithManualEffortFive}{33}
\newcommand{\cPapersWithManualEffortFive}{ICPC}
\newcommand{\pPapersWithManualEffortSix}{33}
\newcommand{\cPapersWithManualEffortSix}{MSR}
\newcommand{\pPapersWithManualEffortSeven}{37}
\newcommand{\cPapersWithManualEffortSeven}{WCRE}

\newcommand{\numPapersWithReposCSMR}{3}
\newcommand{\numPapersWithReposESEM}{0}
\newcommand{\numPapersWithReposICPC}{1}
\newcommand{\numPapersWithReposICSM}{2}
\newcommand{\numPapersWithReposMSR}{2}
\newcommand{\numPapersWithReposSCAM}{1}
\newcommand{\numPapersWithReposWCRE}{3}

\newcommand{\numPapersWithPrevWorkCSMR}{2}
\newcommand{\numPapersWithPrevWorkESEM}{3}
\newcommand{\numPapersWithPrevWorkICPC}{2}
\newcommand{\numPapersWithPrevWorkICSM}{1}
\newcommand{\numPapersWithPrevWorkMSR}{1}
\newcommand{\numPapersWithPrevWorkSCAM}{2}
\newcommand{\numPapersWithPrevWorkWCRE}{2}

\newcommand{\pPapersWithProjectCorpusOne}{58}
\newcommand{\cPapersWithProjectCorpusOne}{ESEM}
\newcommand{\pPapersWithProjectCorpusTwo}{81}
\newcommand{\cPapersWithProjectCorpusTwo}{ICPC}
\newcommand{\pPapersWithProjectCorpusThree}{81}
\newcommand{\cPapersWithProjectCorpusThree}{WCRE}
\newcommand{\pPapersWithProjectCorpusFour}{84}
\newcommand{\cPapersWithProjectCorpusFour}{SCAM}
\newcommand{\pPapersWithProjectCorpusFive}{87}
\newcommand{\cPapersWithProjectCorpusFive}{CSMR}
\newcommand{\pPapersWithProjectCorpusSix}{89}
\newcommand{\cPapersWithProjectCorpusSix}{MSR}
\newcommand{\pPapersWithProjectCorpusSeven}{94}
\newcommand{\cPapersWithProjectCorpusSeven}{ICSM}

\newcommand{\pMaxProjInCSMR}{23}

\newcommand{\pMaxProjInESEM}{13}

\newcommand{\pMaxProjInICPC}{24}

\newcommand{\pMaxProjInICSM}{14}

\newcommand{\pMaxProjInMSR}{28}

\newcommand{\pMaxProjInSCAM}{11}

\newcommand{\pMaxProjInWCRE}{11}

\newcommand{\pPapersWithNoQualitySignsOne}{14}
\newcommand{\cPapersWithNoQualitySignsOne}{ICPC}
\newcommand{\pPapersWithNoQualitySignsTwo}{17}
\newcommand{\cPapersWithNoQualitySignsTwo}{MSR}
\newcommand{\pPapersWithNoQualitySignsThree}{19}
\newcommand{\cPapersWithNoQualitySignsThree}{WCRE}
\newcommand{\pPapersWithNoQualitySignsFour}{21}
\newcommand{\cPapersWithNoQualitySignsFour}{ESEM}
\newcommand{\pPapersWithNoQualitySignsFive}{22}
\newcommand{\cPapersWithNoQualitySignsFive}{ICSM}
\newcommand{\pPapersWithNoQualitySignsSix}{27}
\newcommand{\cPapersWithNoQualitySignsSix}{CSMR}
\newcommand{\pPapersWithNoQualitySignsSeven}{53}
\newcommand{\cPapersWithNoQualitySignsSeven}{SCAM}

\newcommand{\pPapersWithOnlyThreatsOne}{4}
\newcommand{\cPapersWithOnlyThreatsOne}{ESEM}
\newcommand{\pPapersWithOnlyThreatsTwo}{11}
\newcommand{\cPapersWithOnlyThreatsTwo}{MSR}
\newcommand{\pPapersWithOnlyThreatsThree}{11}
\newcommand{\cPapersWithOnlyThreatsThree}{SCAM}
\newcommand{\pPapersWithOnlyThreatsFour}{15}
\newcommand{\cPapersWithOnlyThreatsFour}{WCRE}
\newcommand{\pPapersWithOnlyThreatsFive}{17}
\newcommand{\cPapersWithOnlyThreatsFive}{CSMR}
\newcommand{\pPapersWithOnlyThreatsSix}{19}
\newcommand{\cPapersWithOnlyThreatsSix}{ICPC}
\newcommand{\pPapersWithOnlyThreatsSeven}{31}
\newcommand{\cPapersWithOnlyThreatsSeven}{ICSM}

\newcommand{\pPapersWithRQsAndThreatsOne}{8}
\newcommand{\cPapersWithRQsAndThreatsOne}{ICSM}
\newcommand{\pPapersWithRQsAndThreatsTwo}{11}
\newcommand{\cPapersWithRQsAndThreatsTwo}{WCRE}
\newcommand{\pPapersWithRQsAndThreatsThree}{16}
\newcommand{\cPapersWithRQsAndThreatsThree}{SCAM}
\newcommand{\pPapersWithRQsAndThreatsFour}{20}
\newcommand{\cPapersWithRQsAndThreatsFour}{CSMR}
\newcommand{\pPapersWithRQsAndThreatsFive}{22}
\newcommand{\cPapersWithRQsAndThreatsFive}{MSR}
\newcommand{\pPapersWithRQsAndThreatsSix}{24}
\newcommand{\cPapersWithRQsAndThreatsSix}{ICPC}
\newcommand{\pPapersWithRQsAndThreatsSeven}{42}
\newcommand{\cPapersWithRQsAndThreatsSeven}{ESEM}

\newcommand{\pPapersWithThreeTopQualityTypesWCRE}{44}

\newcommand{\pPapersWithIntroducedToolsOne}{4}
\newcommand{\cPapersWithIntroducedToolsOne}{ESEM}
\newcommand{\pPapersWithIntroducedToolsTwo}{11}
\newcommand{\cPapersWithIntroducedToolsTwo}{MSR}
\newcommand{\pPapersWithIntroducedToolsThree}{19}
\newcommand{\cPapersWithIntroducedToolsThree}{ICPC}
\newcommand{\pPapersWithIntroducedToolsFour}{21}
\newcommand{\cPapersWithIntroducedToolsFour}{SCAM}
\newcommand{\pPapersWithIntroducedToolsFive}{30}
\newcommand{\cPapersWithIntroducedToolsFive}{CSMR}
\newcommand{\pPapersWithIntroducedToolsSix}{37}
\newcommand{\cPapersWithIntroducedToolsSix}{WCRE}
\newcommand{\pPapersWithIntroducedToolsSeven}{44}
\newcommand{\cPapersWithIntroducedToolsSeven}{ICSM}

\newcommand{\pPapersWithNonameToolsOne}{5}
\newcommand{\cPapersWithNonameToolsOne}{ICPC}
\newcommand{\pPapersWithNonameToolsTwo}{17}
\newcommand{\cPapersWithNonameToolsTwo}{ESEM}
\newcommand{\pPapersWithNonameToolsThree}{20}
\newcommand{\cPapersWithNonameToolsThree}{CSMR}
\newcommand{\pPapersWithNonameToolsFour}{33}
\newcommand{\cPapersWithNonameToolsFour}{ICSM}
\newcommand{\pPapersWithNonameToolsFive}{33}
\newcommand{\cPapersWithNonameToolsFive}{MSR}
\newcommand{\pPapersWithNonameToolsSix}{33}
\newcommand{\cPapersWithNonameToolsSix}{WCRE}
\newcommand{\pPapersWithNonameToolsSeven}{42}
\newcommand{\cPapersWithNonameToolsSeven}{SCAM}

\newcommand{\pMaxWexpInCSMR}{3}

\newcommand{\pMaxWexpInESEM}{25}

\newcommand{\pMaxWexpInICPC}{19}

\newcommand{\pMaxWexpInICSM}{8}

\newcommand{\pMaxWexpInMSR}{0}

\newcommand{\pMaxWexpInSCAM}{5}

\newcommand{\pMaxWexpInWCRE}{7}

\newcommand{\pPapersWithSelfTypesOne}{37}
\newcommand{\cPapersWithSelfTypesOne}{SCAM}
\newcommand{\pPapersWithSelfTypesTwo}{61}
\newcommand{\cPapersWithSelfTypesTwo}{MSR}
\newcommand{\pPapersWithSelfTypesThree}{80}
\newcommand{\cPapersWithSelfTypesThree}{CSMR}
\newcommand{\pPapersWithSelfTypesFour}{81}
\newcommand{\cPapersWithSelfTypesFour}{WCRE}
\newcommand{\pPapersWithSelfTypesFive}{86}
\newcommand{\cPapersWithSelfTypesFive}{ICPC}
\newcommand{\pPapersWithSelfTypesSix}{88}
\newcommand{\cPapersWithSelfTypesSix}{ESEM}
\newcommand{\pPapersWithSelfTypesSeven}{94}
\newcommand{\cPapersWithSelfTypesSeven}{ICSM}

\newcommand{\conferences}{CSMR, ESEM, ICPC, ICSM, MSR, SCAM, and WCRE}
\newcommand{\numPapersSurveyed}{175}

\newcommand{\moreThanOneSelfType}{24}
\newcommand{\noSelfTypeNum}{36}

\newcommand{\numPapersWithReplication}{8}
\newcommand{\numPapersWithSelfReplication}{3}

\newcommand{\papersWithMoreThanOneCorpus}{28}
\newcommand{\totalCorpora}{198}
\newcommand{\corporaWithProjects}{168}

\newcommand{\papersWithCorporaWithProjects}{145}

\newcommand{\numPapersWithLargeCorpora}{8}

\newcommand{\numLangJava}{106}
\newcommand{\numLangCLike}{50}
\newcommand{\langCLike}{C, C++, C\#}

\newcommand{\numOnlyOpenSource}{128}
\newcommand{\numOnlyClosedSource}{12}
\newcommand{\numOnlySelfWritten}{9}

\newcommand{\numSrc}{125}
\newcommand{\numBin}{15}

\newcommand{\numPapersWithCorporaWithEclipseGrouped}{22}

\newcommand{\numCorporaWithBugReports}{21}
\newcommand{\numCorporaWithDefects}{16}
\newcommand{\numCorporaWithTests}{10}
\newcommand{\numCorporaWithTraces}{5}

\newcommand{\numPapersWithDatasets}{20}
\newcommand{\numPapersWithPrevWork}{13}
\newcommand{\numPapersWithRepos}{12}

\newcommand{\numPapersWithSources}{43}
\newcommand{\pPapersWithSources}{25}

\newcommand{\papersWithEvolutionMeasures}{52}
\newcommand{\papersWithVersions}{21}
\newcommand{\papersWithRevisions}{11}
\newcommand{\papersWithReleases}{11}
\newcommand{\papersWithCommits}{10}

\newcommand{\papersMentionVCS}{23}
\newcommand{\numCVS}{11}
\newcommand{\numSVN}{11}
\newcommand{\numGit}{4}
\newcommand{\numMerc}{2}

\newcommand{\numPapersWithTimeSpan}{46}
\newcommand{\numPapersWithYears}{36}

\newcommand{\numPapersWithReqEcosystem}{37}
\newcommand{\numPapersWithReqSize}{25}
\newcommand{\numPapersWithReqPreviousWork}{23}
\newcommand{\numPapersWithReqLang}{22}
\newcommand{\numPapersWithReqDomain}{14}
\newcommand{\numPapersWithReqUsage}{14}
\newcommand{\numPapersWithReqHistory}{15}
\newcommand{\numPapersWithReqDependencies}{11}
\newcommand{\numPapersWithReqTests}{10}

\newcommand{\numPapersWithModifications}{20}
\newcommand{\numPapersWithTestsRun}{15}
\newcommand{\numPapersWithRun}{10}

\newcommand{\numPapersWithContentsFiltering}{6}

\newcommand{\numCorporaBenchmarkOracle}{6}
\newcommand{\numCorporaTraining}{6}
\newcommand{\numCorporaEvaluation}{5}

\newcommand{\numCorporaTesting}{4}
\newcommand{\numCorporaQuality}{4}

\newcommand{\numCorporaManualNone}{120}

\newcommand{\numCorporaManualSomeGrouped}{33}
\newcommand{\numCorporaManualAll}{10}

\newcommand{\numExperiments}{22}
\newcommand{\numExperimentsWithPilotStudy}{6}
\newcommand{\numExperimentsWithCompensation}{6}
\newcommand{\numExperimentsWithCorpora}{21}
\newcommand{\numExperimentsWithProjectBasedCorpora}{17}
\newcommand{\numQsInExperiments}{20}
\newcommand{\numQsNotInExperiments}{16}
\newcommand{\numExperimentsWithQ}{10}

\newcommand{\numQuestionnaires}{36}
\newcommand{\numQuestionnairesWithCorpus}{6}
\newcommand{\numQuestionnairesWithPilotStudy}{5}
\newcommand{\numQuestionnairesWithRatio}{6}
\newcommand{\numQuestionnairesPretest}{6}
\newcommand{\numQuestionnairesPosttest}{9}

\newcommand{\numLitSurverys}{6}

\newcommand{\pSelectedPapersInLitSurverysOnAverage}{1.6}

\newcommand{\numPapersWithComparisonsWOProjectBasedCorpus}{5}

\newcommand{\numPapersWithToolsNoIntroducedNoNoname}{126}

\newcommand{\numPapersWithNonameTools}{46}
\newcommand{\numPapersWithNameTools}{46}
\newcommand{\numPapersToolEclipse}{25}

\newcommand{\numPapersToolR}{16}
\newcommand{\numPapersToolUnderstand}{6}
\newcommand{\numPapersToolWeka}{6}
\newcommand{\numPapersToolCCFinder}{6}
\newcommand{\numPapersToolConqat}{4}
\newcommand{\numPapersToolMallet}{4}
\newcommand{\numPapersToolChangedistiller}{3}
\newcommand{\numPapersToolCodesurfer}{3}
\newcommand{\numPapersToolEvolizer}{3}
\newcommand{\numPapersToolRapidminer}{3}
\newcommand{\numPapersToolRecoder}{3}
\newcommand{\numToolsUsedInTwoPapers}{19}

\newcommand{\numPapersWithDefinitions}{25}
\newcommand{\numPapersWithResearchQuestions}{83}
\newcommand{\numPapersWithGQM}{22}
\newcommand{\numPapersWithHypotheses}{23}
\newcommand{\numPapersWithThreatsSection}{111}
\newcommand{\numPapersWithWohlinThreats}{75}
\newcommand{\numPapersWithExternal}{73}
\newcommand{\numPapersWithInternal}{59}
\newcommand{\numPapersWithConstruct}{53}
\newcommand{\numPapersWithConclusion}{26}

\newcommand{\numPapersWithoutAnything}{42}
\newcommand{\numPapersWithRQsAndThreats}{34}
\newcommand{\numPapersWithOnlyThreats}{29}

\newcommand{\pPapersThreeTopCombinations}{60}

\newcommand{\numPapersWithSomeValidation}{88}
\newcommand{\numPapersWithManualValidation}{50}
\newcommand{\numPapersWithValidationAgainst}{27}
\newcommand{\numPapersWithCrossValidation}{8}

\newcommand{\numPapersWithProvidedReplicationPackage}{61}
\newcommand{\numReplicationLinksLeadingToNothing}{6}
\newcommand{\numReplicationLinksWrong}{3}
\newcommand{\numProvidedTools}{25}
\newcommand{\numProvidedCorpus}{15}
\newcommand{\numProvidedRawData}{14}

\newcommand{\numProvidedDescOfCorpus}{6}

\newcommand{\numPapersWithVersions}{67}
\newcommand{\numPapersWithNoVersions}{21}
\newcommand{\numPapersWithTimePeriods}{26}

\newcommand{\numPapersReproducible}{29}

\newcommand{\numPapersHighGrouped}{81}
\newcommand{\numPapersModerate}{78}

\newcommand{\numPapersLowGrouped}{16}

\newcommand{\numPapersNonSearchablePDFs}{5}

\begin{document}

%\title{Contemporary Empirical Research\\ in Software Engineering}
%\title{A Literature Survey on Corpora in Empirical Software Engineering}
\title{
%A Literature Survey on\\Empirical Software Engineering Research
%A Survey of Software Engineering Research:\\ Identifying Used Empirical Evidence
A Literature Survey on \\Empirical Evidence in Software Engineering
}

\author{
\IEEEauthorblockN{Ekaterina Pek}
\IEEEauthorblockA{
Software Languages Team, ADAPT Lab\\
Universit\"at Koblenz-Landau, Germany
}
\and
\IEEEauthorblockN{Ralf L{\"a}mmel}
\IEEEauthorblockA{
Software Languages Team\\
Universit\"at Koblenz-Landau, Germany
}
}

\maketitle

\begin{abstract}
\begin{comment}
  This paper describes the methodology and the results of a literature
  survey on empirical evidence use in Software Engineering. To
  this end, we analyzed the latest proceedings of seven relevant
  conferences. %: \conferences{}. 
  For each paper in the resulting
  collection, we extracted information about the general structure of
  the paper, of the empirical evidence used (contents, sources,
  requirements, etc.); we also assessed reproducibility of the
  research reported in the papers as well as our confidence in our analysis results. The
  results of the survey are discussed. One finding is that despite the
  common use of source code of open-source Java projects, there are no
  frequently used projects across all the papers, but in some
  conferences we can detect recurrences. Overall, the survey feeds
  into a quantitative basis for discussing the current state of
  empirical research in software engineering, for example, concerning
  reproducibility of studies.
\end{comment}
\emph{Context:} Software Engineering research makes use of collections of
software artifacts (corpora) to derive empirical evidence from.
%\emph{Contribution:} 
\emph{Goal:} To improve quality and reproducibility of research, we
need to understand the characteristics of used corpora.
\emph{Method:} For that, we perform a literature survey using grounded
theory. We analyze the latest proceedings of seven relevant
conferences. \emph{Results:} While almost all papers use corpora of
some kind with the common case of collections of source code of
open-source Java projects, there are no frequently used projects or
corpora across all the papers. For some conferences we can detect
recurrences. We discover several forms of requirements and applied
tunings for corpora which indicate more specific needs of research
efforts. \emph{Conclusion:} Our survey feeds into a quantitative basis
for discussing the current state of empirical research in software
engineering, thereby enabling ultimately improvement of research
quality specifically in terms of use (and reuse) of empirical
evidence.
\end{abstract}

%-------------------------------------------------------------------------

%-------------------------------------------------------------------------

\section{Introduction}
\label{S:intro}

%-------------------------------------------------------------------------

This is a survey on software engineering research with focus
on the use of collections of software artifacts (corpora) to derive
empirical evidence from. Such focus on corpora was triggered by our
own research on specifically software reverse/re-engineering and
program comprehension, e.g., studies on API or language
usage~\cite{LaemmelKR05,LaemmelPS11,LaemmelLPV11,LammelP13}---with the
common use of corpora for validation in the broader sense. The survey
applies to conferences that fit with this context. One can observe a
diversity of involved methodologies and characteristics of the
collections of empirical evidence as they are leveraged in
SE research. Thus, we embarked on the present literature survey with
the following central research questions:

\begin{enumerate}[I]
\item How often do Software Engineering papers use \emph{corpora}---collections of empirical evidence?
\item What is the nature and characteristics of the used corpora?
\item Does common contents occur in the used corpora?
\end{enumerate}
 
%and we leveraged grounded theory to discover the underlying coding scheme.

For this, we collected and analyzed the latest proceedings\footnote{As
  they are available from the DBLP bibliography service,
  \url{http://dblp.uni-trier.de/}} of the following conferences:
European Conference on Software Maintenance and Reengineering (CSMR),
International Symposium on Empirical Software Engineering and
Measurement (ESEM), International Conference on Program Comprehension
(ICPC), International Conference on Software Maintenance (ICSM),
Working Conference on Mining Software Repositories (MSR), Working
Conference on Source Code Analysis and Manipulation (SCAM), Working
Conference on Reverse Engineering (WCRE). We choose these conferences
because i) they cover software engineering topics that, based on our
experience, we expect to make use of empirical evidence; ii) they
cover ground related to our expertise and research focus on software
reverse/re-engineering and program comprehension with ESEM as notable
addition for broader coverage of empirical software engineering
research; iii) the conferences are of comparable size. In our survey,
we use only long papers. We choose to analyze only conference
proceedings, because while journal articles may adhere to the best
practices, conference proceedings arguably contain the most common
practices of research in the community---and we are interested in the
latter.

%-------------------------------------------------------------------------
%-------------------------------------------------------------------------

%\begin{sidewaystable*}
\begin{table*}

\begin{threeparttable}

\begin{center}
\caption{Literature surveys on Software Engineering research}
\label{T:rel}
\begin{tabular}{@{}v{5.5em}*{4}{r}v{5em}*{3}{r}v{10em}v{20em}@{}{r}@{}}
\toprule

Name & Ref & Year & \multicolumn{2}{c}{\# used} & Period & \multicolumn{3}{c}{\# papers} & Focus & Coding schema & \\
\cmidrule(lr){4-5} \cmidrule(lr){7-9}
 & & & j & c & & total & sel. & rel. & & & \\
\midrule

Glass et al. 
  & \cite{GlassVR02}
  & 2002
  & 6
  & 0
  & 1995--1999
  & ---
  & 369
  & 369
%  & ---
  & Characteristics of SE research
%  & RQ
  & Topics, research approaches and methods, theoretical basis, level of analysis
  &
\\
\addlinespace

Sj{\o}berg et al. 
  & \cite{SjobergEtAl05}
  & 2005
  & 9
  & 3
  & 1993--2002
  & 5453
  & 103
  & 103
%  & ---
  & Controlled experiments
%  & ---
  & Extent, topic, subjects, task and environment, replication, internal and external validity
  &
\\
\addlinespace

Zannier et al. 
  & \cite{ZannierEtAl06}
  & 2006
  & 0
  & 1
  & 1975--2005
  & 1227
  & 63
  & 44
%  & Quasi-random experiment
  & Empirical evaluation: quantity and soundness 
%  & H
  & Study type, sampling type, target and used population, evaluaton type, proper use of analysis, usage of hypotheses
  &
\\
\addlinespace

Kitchenham et al. 
  & \cite{KitchenhamEtAl09}
  & 2009
  & 10
  & 3
  & 2004--2007
  & 2506
  & 33
  & 19
%  & Systematic review
  & Systematic reviews
%  & RQ
  & Inclusion and exclusion criteria, coverage, quality/validity assessment, description of the basic data
  &
\\
\addlinespace

\textbf{Our study}
  & 
  & \textbf{2013}
  & \textbf{0}
  & \textbf{7}
  & \textbf{2011/2012}
  & \textbf{227}
  & \textbf{175}
  & \textbf{175}
%  & Grounded theory
  & \textbf{Empirical evidence}
%  & ---
  & \textbf{Emerged classification}
  &
\\
\addlinespace

\bottomrule
\end{tabular}

\begin{tablenotes}
\item {\scriptsize 
Legend: 
\emph{j} and \emph{c} stand for \emph{journals} and \emph{conferences}; 
\emph{sel.} and \emph{rel.} stand for \emph{selected} and \emph{relevant}.}
\end{tablenotes}

\end{center}

\end{threeparttable}

%\end{sidewaystable*}
\end{table*}

%-------------------------------------------------------------------------

%-------------------------------------------------------------------------

SE research has been surveyed before; see Table~\ref{T:rel} for a
summary. The cited surveys focus on specific forms or characteristics
of SE research to be analyzed with a predefined schema. For instance,
Kitchenham et al.\ surveyed SE journals and conferences to find out
adoption rate of systematic literature
reviews~\cite{KitchenhamEtAl09}. Similarly, Sj{\o}berg et al.\ sought
to find and analyze existing controlled experiments in SE
research~\cite{SjobergEtAl05}. By contrast, we (first to our
knowledge) seek to discover whatever empirical evidence is used to
facilitate SE research and we allow our coding schema to emerge from
the data. We follow the idea of Grounded Theory (GT) as understood by
Glaser~\cite{Glaser08}~(on the difference between Straussarian and
Glaserian versions see~\cite{AdolphEtAl08}).

\begin{comment}
% Not so clear from the table
Usually, literature surveys are restricted to a specific context and
therefore filter the papers they collect so that only those papers are
analyzed that are considered relevant. 
\end{comment}

%\paragraph*{Road-map of the paper} 
The paper is organized as follows: \S\ref{S:meth} describes the
methodology underlying this literature survey. \S\ref{S:res} presents
the results of the survey. \S\ref{S:rel} discusses related
surveys. \S\ref{S:threats} identifies threats to
validity. \S\ref{S:concl} concludes the paper.

%-------------------------------------------------------------------------

%-------------------------------------------------------------------------

\section{Methodology}
\label{S:meth}

%-------------------------------------------------------------------------

Empirical research is usually perceived as taking one of the forms:
controlled and quasi-experiments, exploratory and confirmatory case
studies, survey, ethnography, and action
research~\cite{SjobergEtAl07,EasterbrookEtAl08}. In a broader sense,
empirical research also includes any research based on collected
evidence---quoting from~\cite{SjobergEtAl07}: ``Empirical research
seeks to explore, describe, predict, and explain natural, social, or
cognitive phenomena \emph{by using evidence based on observation} or
experience. It involves obtaining and interpreting evidence by, e.g.,
experimentation, systematic observation, interviews or surveys, or
\emph{by the careful examination of documents or artifacts} [emphasis
added].''

Since Software Engineering is a practical area of Computer Science, it
is logical to expect that most of the SE research is evidence-based,
i.e., empirical de facto and in the present study, we submit to
substantiate this expectation. We believe that a bottom-up approach
of observing what exists and discovering methodology as well as
definitions of forms of research complements the prominent top-down
approach, when a methodology is derived from theoretical
considerations or by borrowing from other sciences (medicine,
sociology, psychology).

This survey is particularly concerned with (collections of) empirical
evidence. Thus, the following questions guide the research:

\begin{enumerate}[I]
\item How often do Software Engineering papers use \emph{corpora}---collections of empirical evidence?
\item What is the nature and characteristics of the used corpora? 
\item Does common contents occur in the used corpora?
\end{enumerate}

%-------------------------------------------------------------------------

%-------------------------------------------------------------------------

\begin{table}
\begin{center}
\caption{Conferences used in the survey}
\label{T:conf}
\begin{tabular}{@{}{r}{l}*{2}{r}@{}}
\toprule
Year & Conference & \multicolumn{2}{c}{\# papers} \\
\cmidrule(lr){3-4}
 & & total & long \\
\midrule

2012 & CSMR & 30 & 30 \\
2012 & ESEM & 43 & 24 \\
2012 & ICPC & 23 & 21 \\
2011 & ICSM & 36 & 36 \\
2012 & MSR & 29 & 18 \\
2011 & SCAM & 19 & 19 \\
2011 & WCRE & 47 & 27 \\

\midrule

& Total & 227 & \numPapersSurveyed{} \\

\bottomrule
\end{tabular}
\end{center}
\end{table}

%-------------------------------------------------------------------------

%-------------------------------------------------------------------------

For that, we collected the papers from the latest edition of seven SE
conferences: \conferences{} (see~Table~\ref{T:conf} for details).  We
used DBLP pages of conferences to identify long papers and downloaded
them from digital libraries.  

We then proceeded to read the papers to
perform coding. From a previously done, smaller and more specific
literature survey~\cite{FavreEtAl11} and a pilot study for the present
survey, we had some basic understanding of the parts of the scheme to
emerge. During the first pass of coding, we started with the empty
scheme and completed it eventually to arrive at the current scheme, as
described below. During the second pass, we compared profiles of coded
papers against the latest version of the scheme, we went through the
papers again and filled in the missing details.

While we were interested primarily in characteristics of used
empirical evidence (specifically corpora), we also extracted
additional information about research reported in the papers: used
tools, signs of rigorousness/quality, etc.  We put the collected
information in several groups:

\subsubsection{Corpora} 
We captured what was used as study objects (e.g., projects), what are their characteristics (e.g.,
language, open- vs. closed-source, code form), what are the requirements to the
study objects, do they come from a specific source (e.g., established dataset
or online repository), were they observed over a time (e.g., versions or
revisions), 
what is the nature of preparation of the corpus.

\subsubsection{Forms of empirical research}
During coding, several structural forms evolved that we used for capturing
information conveyed in papers: experiments, questionnaires, literature
surveys, and comparisons. Some relationships between forms and corpora
usage also emerged.

\subsubsection{Self-classification}
For each paper we captured what words authors use to describe their effort: e.g., case study, experiment.

\subsubsection{Tools}
We collected mentions of existing tools (e.g., Eclipse, R, Weka) that were used as well as of introduced
tools that were presented in the papers. (In many cases, these tools
are used to analyze or to otherwise process corpora.)

\subsubsection{Structural signs of rigorousness/quality} 
We paid attention to the following aspects of the study presentation:
Do authors use research questions? Null hypotheses? 
Is there a section on definitions and terms?
Is validation mentioned?
Is there a ``Threats to validity'' section? 
Are threats addressed in any structured way?
%Which threats are mentioned? 

\subsubsection{Reproducibility} We tried to understand in each case,
if a study can be reproduced. (Obviously, the use of corpora affects
the definition of reproducibility.) We paid attention to the following signs: 
Are all details provided for a possible study replication (i.e., versions of used projects, time periods, etc.)?
Do authors provide any material used in the paper, e.g., on a supplementary website?
Altogether, would it be possible to reproduce the study?

\subsubsection{Assessment} Finally, we characterized the process of
coding: how easy it was to extract information and how confident we are in the
result.

We did the pilot survey in September-October~2012.
After that, we adjusted our methodology (e.g., instead of filtering papers
based on their abstract, we decided to survey all the papers) and proceeded to
perform the current study in November~2012-January~2013.
We used Python and Bash scripts, Google Refine
tool\footnote{\url{http://code.google.com/p/google-refine/}}, 
and R project\footnote{\url{http://www.r-project.org/}} to process the data. 
We provide online the list of the papers and results of coding\footnote{\url{http://softlang.uni-koblenz.de/empsurvey}}. 

%-------------------------------------------------------------------------

%-------------------------------------------------------------------------

\section{Results}
\label{S:res}

%-------------------------------------------------------------------------

In this section, we present the results of our study.
We group them similarly to the description provided in Section~\ref{S:meth}:
details about detected corpora, emerged forms of empirical research,
used or introduced tools, signs of rigorousness/quality of research,
reproducibility of the studies, and, finally, assessment of our effort. 
When we use the phrase \emph{``on the average''}, we imply the median of the appropriate distribution.

\punchline{
Next to the numbers, we provide framed highlights.

We use formula ``X out of Y papers'' to provide feeling for the numbers. E.g.,
``one out of three papers'' means that in every three surveyed papers there is one that has the discussed characteristic.

We also provide conference-wise percentage of found characteristics.
The table below illustrates the format on an artificial example:
conferences are listed from left to right as the percentage increases. 
Percentage is always given relative to the total number of the long papers in the conference.
Where appropriate, below the percentage appear names of the most popular projects, requirements, tunings within the conferences.
When more than one name is given, each of them appear with the specified frequency.

\centering
{\footnotesize
\begin{tabular}{@{}{r}@{\hskip 0.15in}{r}@{\hskip 0.15in}{r}@{\hskip 0.15in}{r}@{\hskip 0.15in}{r}@{\hskip 0.15in}{r}@{\hskip 0.15in}{r}@{}}
\addlinespace
\addlinespace
\multicolumn{7}{c}{\textbf{Artificial example}} \\
\toprule
CSMR & ESEM & ICPC & ICSM & MSR & SCAM & WCRE \\
1\,\% &
2\,\% &
3\,\% &
4\,\% &
5\,\% &
6\,\% &
7\,\% \\
\bottomrule
\end{tabular}
}

}

%-------------------------------------------------------------------------

%-------------------------------------------------------------------------
%\input{sources.tex}
%-------------------------------------------------------------------------

\subsection{Corpora}
\label{S:corpora}

%-------------------------------------------------------------------------

\subsubsection{Usage}

% used_for = ""

We marked a paper as containing a corpus when the paper mentioned a collection of software artifacts used for deriving empirical evidence.
Altogether, we have found \totalCorpora{} corpora used in \freqCorp{} papers out of \numPapersSurveyed{} surveyed papers.

In \papersWithMoreThanOneCorpus{} cases, we decided that a paper contains more than one corpus. 
We did so consistently, when we met at least two of the
following motivations mentioned in the paper when describing the purpose of
collected empirical evidence:
for benchmark or oracle (\numCorporaBenchmarkOracle{} corpora), for training (\numCorporaTraining{} corpora),
for evaluation (\numCorporaEvaluation{} corpora), for investigation (\numCorporaEvaluation{} corpora), 
for testing (\numCorporaTesting{} corpora), for investigating quality like accuracy or scalability (\numCorporaQuality{} corpora).

We have found that \corporaWithProjects{} corpora (used in \papersWithCorporaWithProjects{}~papers), 
consist of projects (systems, software); 
in other cases, corpora consist of another kind of study object: image, trace, feature, web log, etc.
Till the end of the current subsection~(\ref{S:corpora}), we restrict ourselves
to the corpora consisting of projects and call them project-based corpora. 

\punchline{
Almost all papers use a corpus of some sort. One out of six papers has more than one corpus.
Most of the corpora consist of projects.

%Across conferences, ICSM shows the highest usage of project-based corpora~(\pPapersWithProjectCorpusICSM{}\,\%~papers);
%ESEM---the lowest~(\pPapersWithProjectCorpusESEM{}\,\%~papers).

%\begin{minipage}{\linewidth}
\centering
{\footnotesize
\begin{tabular}{@{}{r}@{\hskip 0.15in}{r}@{\hskip 0.15in}{r}@{\hskip 0.15in}{r}@{\hskip 0.15in}{r}@{\hskip 0.15in}{r}@{\hskip 0.15in}{r}@{}}
\addlinespace
\addlinespace
\multicolumn{7}{c}{\textbf{Project-based corpora usage}} \\
\toprule
\cPapersWithProjectCorpusOne{} &
\cPapersWithProjectCorpusTwo{} &
\cPapersWithProjectCorpusThree{} &
\cPapersWithProjectCorpusFour{} &
\cPapersWithProjectCorpusFive{} &
\cPapersWithProjectCorpusSix{} &
\cPapersWithProjectCorpusSeven{} 
\\

\pPapersWithProjectCorpusOne{}\,\% & 
\pPapersWithProjectCorpusTwo{}\,\% & 
\pPapersWithProjectCorpusThree{}\,\% & 
\pPapersWithProjectCorpusFour{}\,\% & 
\pPapersWithProjectCorpusFive{}\,\% & 
\pPapersWithProjectCorpusSix{}\,\% & 
\pPapersWithProjectCorpusSeven{}\,\% 
\\ 
\bottomrule
\end{tabular}
}
%\end{minipage}

}

%-------------------------------------------------------------------------

\subsubsection{Contents}
We identified the following common characteristics of project-based corpora.
%Further in this section, we provide more details for the project-based corpora. 

%-------------------------------------------------------------------------

%      num_projects =
\textbf{Size.}
Half of the corpora, \existsLessOrEqThreeNumProjects{} cases, have three or
less projects (of them, \existsEqualOneNumProjects{} corpora
consist of only one project). There are \existsMoreThanTenNumProjects{}
corpora that contain more than \valueMoreThanTenNumProjects{} projects. 
%\kate{Examples the large corpora? I.e., with more than 10 projects.} 
We detected large corpora (with more than 100 projects) in~\numPapersWithLargeCorpora{}~papers---one of them introducing an established
dataset itself.

%-------------------------------------------------------------------------

%      languages = ""
\textbf{Languages.}  Most of the corpora are monolingual
(\existsEqualOneNumLang{} cases); most of the remaining ones are
bilingual (\existsEqualTwoNumLang{} cases).  As for the software
language, \numLangJava{} corpora contain projects written in Java,
while C-like languages are used in \numLangCLike{} corpora (in C-like
languages we include \langCLike{}).

%-------------------------------------------------------------------------

%      code_form = ""
\textbf{Code form.}
In \numSrc{} cases, corpora consist of source code; in \numBin{} cases---of
binaries. In the rest of the cases, code of the projects is not used, something
else is in focus (developers, requirements, etc.) 

%-------------------------------------------------------------------------

%      open_source = true
%      closed_source = false
\textbf{Access.}
In \numOnlyOpenSource{} cases, corpora consist only of open-source projects; 
in \numOnlyClosedSource{} cases, corpora consist only of projects not available publicly (e.g., industrial software); 
in \numOnlySelfWritten{} cases, corpora are self-written. The
remaining cases mix access forms.

%-------------------------------------------------------------------------

%      projects = ""
\textbf{Projects.}
We collected names of the used projects as they are provided by the papers
(modulo merging of names like Vuze/Azureus\footnote{The project changed its
name in 2008.}).
%-------------------------------------------------------------------------

\begin{wraptable}{l}{3cm}
{\small
\caption{Used projects}
\label{T:projects}
\begin{tabular}{@{}{l}{r}@{}}
\toprule
Project & \# corp \\
\midrule

JHotDraw & 15 \\ 
JEdit & 12 \\ 
Ant & 11 \\ 
ArgoUML & 11 \\ 
Eclipse & 11 \\ 
Firefox & 10 \\ 
Vuze/Azureus & 8 \\ 
Linux kernel & 6 \\ 
Lucene & 6 \\ 
Mozilla & 6 \\ 
Hibernate & 5 \\

\bottomrule
\end{tabular}
}
\end{wraptable}

%-------------------------------------------------------------------------

Table~\ref{T:projects} lists projects frequently used in the corpora.
\begin{comment}
The top of the usage frequency is as follows:
JHotDraw (used in~15~corpora), 
JEdit~(12), 
Ant~(11), 
ArgoUML~(11), 
Eclipse~(11), 
Firefox~(10), 
Vuze/Azureus~(8),
Linux kernel~(6),
Lucene~(6), 
Mozilla~(6), 
Hibernate~(5).
\end{comment}
Eclipse is a complex project, and some corpora
make use of its sub-parts, considering them as projects on their own (e.g., JDT
Search, PDE Build)---counting such cases, there are altogether \numPapersWithCorporaWithEclipseGrouped{} papers making use of Eclipse.

%-------------------------------------------------------------------------

%      units = ""
%      num_units =
\textbf{Units.}
We captured when 
some unit related to the project
was in the focus of the study: a bug report or a UML class diagram---namely,
we would capture the fact when such unit was used to give quantitative
information (e.g., in a table presenting number of bug reports in the project under
investigation). The most popular units turned out to be bug reports, they
are used in~\numCorporaWithBugReports{}~corpora; defects (faults, failures) are
used in~\numCorporaWithDefects{}~corpora; tests---in~\numCorporaWithTests{};
traces---in~\numCorporaWithTraces{}. 

%-------------------------------------------------------------------------

\punchline{
An average project-based corpus consists of source code of three open-source projects, written in Java.
Eclipse or its sub-parts is used in one out of eight papers using project-based corpora.
The projects used in at least five papers are JHotDraw, JEdit, Ant, ArgoUML, Firefox, Vuze/Azerus, Linux kernel, Lucene, Mozilla, and Hibernate.
Within the corpus, bug reports, defects, tests, and traces can be in the focus of the study.

%Most of its usage, Eclipse~(\pMaxProjFromEclipse{}\,\%) and JHotDraw~(\pMaxProjFromJhotdraw{}\,\%) 
%get from \confMaxProjFromEclipse{} papers;
%JEdit~(\pMaxProjFromJedit{}\,\%) and Ant~(\pMaxProjFromAnt{}\,\%)---from \confMaxProjFromJedit{};
%and ArgoUML~(\pMaxProjFromArgouml{}\,\%)---from \confMaxProjFromArgouml{}.

%Across conferences, we can detect favorite projects: 
%e.g., \projMaxProjInCSMR{} is used~in~\pMaxProjInCSMR{}\,\%~CSMR~papers;
%\projMaxProjInICPC{} is used~in~\pMaxProjInICPC{}\,\%~ICPC~papers;
%\projMaxProjInMSR{}, each ---in~\pMaxProjInMSR{}\,\%~MSR~papers.

\centering
{\footnotesize
\begin{tabular}{@{}v{2em}@{\hskip 0.04in}v{2em}@{\hskip 0.04in}v{2em}@{\hskip 0.04in}v{2em}@{\hskip 0.04in}v{2em}@{\hskip 0.04in}v{2em}@{\hskip 0.04in}v{2em}@{}{r}}
\addlinespace
\addlinespace
\multicolumn{7}{c}{\textbf{Popular projects}} \\
\toprule
\multicolumn{1}{c}{WCRE} &
\multicolumn{1}{c}{SCAM} &
\multicolumn{1}{c}{ESEM} &
\multicolumn{1}{c}{ICSM} &
\multicolumn{1}{c}{CSMR} &
\multicolumn{1}{c}{ICPC} &
\multicolumn{1}{c}{MSR} &
\\

\multicolumn{1}{c}{\pMaxProjInWCRE{}\,\%} &
\multicolumn{1}{c}{\pMaxProjInSCAM{}\,\%} &
\multicolumn{1}{c}{\pMaxProjInESEM{}\,\%} &
\multicolumn{1}{c}{\pMaxProjInICSM{}\,\%} &
\multicolumn{1}{c}{\pMaxProjInCSMR{}\,\%} &
\multicolumn{1}{c}{\pMaxProjInICPC{}\,\%} &
\multicolumn{1}{r}{\pMaxProjInMSR{}\,\%} &
\\

\multicolumn{1}{c}{{\scriptsize Eclipse}} &
\multicolumn{1}{c}{{\scriptsize Lynx}} &
\multicolumn{1}{c}{{\scriptsize Eclipse}} &
\multicolumn{1}{c}{{\tiny ArgoUML}} &
\multicolumn{1}{c}{{\scriptsize Eclipse}} &
\multicolumn{1}{c}{{\scriptsize JEdit}} &
\multicolumn{1}{c}{{\scriptsize Firefox}} &
\\

\multicolumn{1}{c}{{\scriptsize JEdit}} &
\multicolumn{1}{c}{{\scriptsize Minix}} &
 &
 &
 &
 &
\multicolumn{1}{c}{{\scriptsize Eclipse}} &
\\

\bottomrule
\end{tabular}
}

}

%-------------------------------------------------------------------------

%      source {
%        type = ""
%        name = ""
%      }

%\input{datasets.tex}
\textbf{Sources.}
When papers clearly state the source of their corpora, we collected such information.

\begin{table}
\begin{threeparttable}
\caption{Online repositories and established datasets}
\label{T:setsrepos}

\begin{tabular}{lr}

\begin{minipage}{.45\linewidth}
\centering

\begin{tabular}{@{}v{7em}@{ }{r}@{}}
\toprule
Repository & \# papers \\
\midrule

SourceForge\tnote{1} & 6 \\
Apache.org\tnote{2} & 3 \\
GitHub\tnote{3} & 3 \\
Android Market\tnote{4} & 2 \\
CodePlex\tnote{5} & 2 \\

\bottomrule
\end{tabular}

\end{minipage}%

&

\begin{minipage}{.45\linewidth}
\centering

\begin{tabular}{@{}{r}@{ }v{6.3em}@{ }{r}@{}}
\toprule
Ref & Dataset & \# papers \\
\midrule

\cite{DoER05} & SIR & 3 \\ 
\cite{msr} & MSR challenge & 2 \\
\cite{Gueheneuc07} & P-MARt & 2 \\
\cite{promise12} & PROMISE & 2 \\
\cite{TemperoEtAl10} & Qualitas & 2 \\

\bottomrule
\end{tabular}

\end{minipage} 
\end{tabular}
\begin{tablenotes}
\item [1] \url{http://sourceforge.net/}
\item [2] \url{http://projects.apache.org/}
\item [3] \url{https://github.com/}
\item [4] Now known as Google Play, \url{https://play.google.com/store}
\item [5] \url{http://www.codeplex.com/}
\end{tablenotes}
\end{threeparttable}

\end{table}

Online repositories used in more than one paper are listed in Table~\ref{T:setsrepos}.
% \reposInMoreThanOnePaper{}.
The rest of detected online repositories are used in only one paper each:
BlackBerry App World\footnote{\url{http://appworld.blackberry.com/webstore}}, 
Google Code\footnote{\url{http://code.google.com/}}, 
Launchpad\footnote{\url{https://launchpad.net/}}, 
ShareJar\footnote{\url{http://www.sharejar.com/}}.

Established datasets used in more than one paper are listed in Table~\ref{T:setsrepos}.
Some of the other datasets that used only in one paper each:
Bug prediction dataset~\cite{DAmbrosEtAl10}, 
CHICKEN Scheme benchmarks~\footnote{\url{https://github.com/mario-goulart/chicken-benchmarks}}, 
CoCoMe\footnote{\url{http://agrausch.informatik.uni-kl.de/CoCoME}}, 
DaCapo~\cite{BlackburnEtAl06a},
FLOSSMetrics\footnote{\url{http://libresoft.es/research/projects/flossmetrics}}, 
iBUGS\footnote{\url{http://www.st.cs.uni-saarland.de/ibugs/}},
SMG2000 benchmark\footnote{\url{https://asc.llnl.gov/computing_resources/purple/archive/benchmarks/smg/}}, 
SourcererDB~\cite{LinsteadEtAl09}, 
TEFSE challenge\footnote{\url{http://www.cs.wm.edu/semeru/tefse2011/Challenge.htm}}.
%UCI source code dataset\footnote{\url{http://www.ics.uci.edu/~lopes/datasets/}}. 
%\kate{Double check UCI}
Table~\ref{T:sources} summarizes the most popular types of sources and their distribution across conferences.

\punchline{
One out of four project-based corpora uses an established dataset, previous work, or online repository
as a source of the projects.
There is no common frequently used dataset or repository. Only
SourceForge shows moderately frequent usage.
%, other than that five different datasets and four repositories are mentioned in two or three papers.

%Across conferences, MSR shows the highest usage of such sources: \pPapersWithSourcesMSR{}\,\%~papers.

\centering
{\footnotesize
\begin{tabular}{@{}{r}@{\hskip 0.15in}{r}@{\hskip 0.15in}{r}@{\hskip 0.15in}{r}@{\hskip 0.15in}{r}@{\hskip 0.15in}{r}@{\hskip 0.15in}{r}@{}}
\addlinespace
\addlinespace
\multicolumn{7}{c}{\textbf{Usage of corpora sources}} \\
\toprule
\cPapersWithSourcesOne{} &
\cPapersWithSourcesTwo{} &
\cPapersWithSourcesThree{} &
\cPapersWithSourcesFour{} &
\cPapersWithSourcesFive{} &
\cPapersWithSourcesSix{} &
\cPapersWithSourcesSeven{}
\\

\pPapersWithSourcesOne{}\,\% &
\pPapersWithSourcesTwo{}\,\% &
\pPapersWithSourcesThree{}\,\% &
\pPapersWithSourcesFour{}\,\% &
\pPapersWithSourcesFive{}\,\% &
\pPapersWithSourcesSix{}\,\% &
\pPapersWithSourcesSeven{}\,\%
\\
\bottomrule
\end{tabular}
}

}

%-------------------------------------------------------------------------
%-------------------------------------------------------------------------
%-------------------------------------------------------------------------

\begin{table*}[t!]
\begin{center}
\caption{Sources of corpora}
\label{T:sources}
\begin{tabular}{@{}{l}*{8}{r}@{}}
\toprule
Type & \multicolumn{8}{c}{\# papers} \\
\cmidrule(lr){2-9}
 & Total & CSMR & ESEM & ICPC & ICSM & MSR & SCAM & WCRE \\
\midrule

Established dataset & 
  \numPapersWithDatasets{} & \numPapersWithDatasetsCSMR{} & 
  \numPapersWithDatasetsESEM{} & \numPapersWithDatasetsICPC{} & 
  \numPapersWithDatasetsICSM{} & \numPapersWithDatasetsMSR{} & 
  \numPapersWithDatasetsSCAM{} & \numPapersWithDatasetsWCRE{} 
\\ 

Previous work & 
  \numPapersWithPrevWork{} & \numPapersWithPrevWorkCSMR{} & 
  \numPapersWithPrevWorkESEM{} & \numPapersWithPrevWorkICPC{} & 
  \numPapersWithPrevWorkICSM{} & \numPapersWithPrevWorkMSR{} & 
  \numPapersWithPrevWorkSCAM{} & \numPapersWithPrevWorkWCRE{} 
\\ 

Online repository & 
  \numPapersWithRepos{} & \numPapersWithReposCSMR{} & 
  \numPapersWithReposESEM{} & \numPapersWithReposICPC{} & 
  \numPapersWithReposICSM{} & \numPapersWithReposMSR{} & 
  \numPapersWithReposSCAM{} & \numPapersWithReposWCRE{} 
\\ 

\midrule

Total & 
  \numPapersWithSources{} & \numPapersWithSourcesCSMR{} & 
  \numPapersWithSourcesESEM{} & \numPapersWithSourcesICPC{} & 
  \numPapersWithSourcesICSM{} & \numPapersWithSourcesMSR{} & 
  \numPapersWithSourcesSCAM{} & \numPapersWithSourcesWCRE{} 
\\

Percentage & 
  \pPapersWithSources{} & \pPapersWithSourcesCSMR{} & 
  \pPapersWithSourcesESEM{} & \pPapersWithSourcesICPC{} & 
  \pPapersWithSourcesICSM{} & \pPapersWithSourcesMSR{} & 
  \pPapersWithSourcesSCAM{} & \pPapersWithSourcesWCRE{} 
\\

\bottomrule
\end{tabular}
\end{center}
\end{table*}

%-------------------------------------------------------------------------

%-------------------------------------------------------------------------

\subsubsection{Evolution}

%      evolution_measures = ""
%      num_evolution_measures =
%      time_span =

We encountered \papersWithEvolutionMeasures{} papers that use evolution of the
projects in their research, meaning that they operate on several versions,
releases, etc. To describe the evolution measure, the following terms were used: 
``version'' (\papersWithVersions{} times),
``revision'' (\papersWithRevisions{}),
``commit'' (\papersWithCommits{}),
``release'' (\papersWithReleases{}).

On the average, papers mentioning commits use \numprint{\medianNumCommits{}} commits; 
papers with revisions---\numprint{\medianNumRevisions{}} revisions;
with versions---\medianNumVersions{} versions;
with releases---\medianNumReleases{} releases.

There are \numPapersWithTimeSpan{} papers that mention a time span of their
study. In \numPapersWithYears{} cases, the unit of the time span is a year and on
the average such papers are concerned with a \medianNumYears{}-year span. 

%      version_control = ""

We found \papersMentionVCS{} papers to mention what version control
system was involved in the study. 
CVS is mentioned \numCVS{} times, SVN---\numSVN{} times, Git and Mercurial---\numGit{} and \numMerc{} times respectively.

\punchline{
One out of three papers with project-based corpora uses evolution aspect in its research.
In half of the cases, large-scale evolution is involved: 
several thousands commits/revisions or ten versions/releases of projects---often spanning several years of a project's lifetime. 

%Across conferences, MSR shows the highest interest in evolution aspect~(\pPapersWithEvolutionMSR{}\,\% papers);
%ICPC and SCAM---the lowest~(\pPapersWithEvolutionICPC{}\,\% and \pPapersWithEvolutionSCAM\,\% papers respectively).

\centering
{\footnotesize
\begin{tabular}{@{}{r}@{\hskip 0.15in}{r}@{\hskip 0.15in}{r}@{\hskip 0.15in}{r}@{\hskip 0.15in}{r}@{\hskip 0.15in}{r}@{\hskip 0.15in}{r}@{}}
\addlinespace
\addlinespace
\multicolumn{7}{c}{\textbf{Evolution usage}} \\
\toprule
\cPapersWithEvolutionOne{} &
\cPapersWithEvolutionTwo{} &
\cPapersWithEvolutionThree{} &
\cPapersWithEvolutionFour{} &
\cPapersWithEvolutionFive{} &
\cPapersWithEvolutionSix{} &
\cPapersWithEvolutionSeven{}
\\

\pPapersWithEvolutionOne{}\,\% &
\pPapersWithEvolutionTwo{}\,\% &
\pPapersWithEvolutionThree{}\,\% &
\pPapersWithEvolutionFour{}\,\% &
\pPapersWithEvolutionFive{}\,\% &
\pPapersWithEvolutionSix{}\,\% &
\pPapersWithEvolutionSeven{}\,\%
\\
\bottomrule
\end{tabular}
}

}

%-------------------------------------------------------------------------

\subsubsection{Requirements}

%      requirements_implicit {
%        req = ""
%      }

We collected requirements to the corpora: explicit as well as
implicit. For instance, an implicit requirement for a bug tracking
system is inferred if the paper uses bug reports of the projects under
investigation. The most popular direction of requirements is the
presence of some `ecosystem' (found in \numPapersWithReqEcosystem{}
papers): existence of bug tracking systems, mailing lists,
documentation (e.g., user manuals).  Another popular requirement,
found in \numPapersWithReqSize{} papers, has to do with the size of
the projects: small, sufficient, large, or of particular size (as
specific as ``medium of the sizes of the ten most popular Sourceforge
projects''), or the need of diversity of sizes.  In
\numPapersWithReqPreviousWork{} papers, it was stated that the used
projects were chosen because they were used in previous work (of the
same or other authors).  Language-related requirement was present in
\numPapersWithReqLang{} papers for a specific language or for the
diversity of languages in a corpus.  In \numPapersWithReqDomain{}
papers, the choice of projects was attributed to either diversity of
application domains or to a specific domain.  Some aspect of the used
projects was mentioned as essential in \numPapersWithReqUsage{}
papers: active or wide-spread usage, popularity, well-known and
established software.  Other popular requirements include presence of
development history (\numPapersWithReqHistory{} papers), dependencies
(\numPapersWithReqDependencies{} papers), or tests
(\numPapersWithReqTests{} papers).

\punchline{
One out of five papers requires the projects of its corpus to have an
ecosystem: a bug tracker, or a mailing list, or some kind of documentation.
Other requirements focus on the size and language of the projects, application domain,
development history, etc.

%Across conferences, we can detect favorite requirements:
%e.g., \reqMaxReqInCSMR{} for CSMR~(in~\pMaxReqInCSMR{}\,\%~papers), WCRE~(\pMaxReqInWCRE{}\,\%~papers), and ESEM~(\pMaxReqInESEM{}\,\%~papers);
%\reqMaxReqInICPC{} for ICPC~(\pMaxReqInICPC{}\,\%~papers) and ICSM~(\pMaxReqInICSM{}\,\%~papers);
%Equally important~(in~\pMaxReqInMSR{}\,\%~papers) for MSR are requirements about \reqMaxReqInMSR{}.

\centering
{\footnotesize
\begin{tabular}{@{}v{2em}@{\hskip 0.04in}v{2em}@{\hskip 0.04in}v{2em}@{\hskip 0.04in}v{2em}@{\hskip 0.04in}v{2em}@{\hskip 0.04in}v{2em}@{\hskip 0.04in}v{2em}@{}{r}}
\addlinespace
\addlinespace
\multicolumn{7}{c}{\textbf{Popular requirements}} \\
\toprule
\multicolumn{1}{c}{SCAM} &
\multicolumn{1}{c}{MSR} &
\multicolumn{1}{c}{ICSM} &
\multicolumn{1}{c}{ICPC} &
\multicolumn{1}{c}{ESEM} &
\multicolumn{1}{c}{WCRE} &
\multicolumn{1}{c}{CSMR} &
\\

\multicolumn{1}{c}{\pMaxProjInSCAM{}\,\%} &
\multicolumn{1}{c}{\pMaxProjInMSR{}\,\%} &
\multicolumn{1}{c}{\pMaxProjInICSM{}\,\%} &
\multicolumn{1}{c}{\pMaxProjInICPC{}\,\%} &
\multicolumn{1}{c}{\pMaxProjInESEM{}\,\%} &
\multicolumn{1}{c}{\pMaxProjInWCRE{}\,\%} &
\multicolumn{1}{r}{\pMaxProjInCSMR{}\,\%} &
\\

\multicolumn{1}{c}{{\scriptsize domain}} &
\multicolumn{1}{c}{{\scriptsize ecosys}} &
\multicolumn{1}{c}{{\scriptsize size}} &
\multicolumn{1}{c}{{\scriptsize size}} &
\multicolumn{1}{c}{{\scriptsize ecosys}} &
\multicolumn{1}{c}{{\scriptsize ecosys}} &
\multicolumn{1}{c}{{\scriptsize ecosys}} &
\\

\multicolumn{1}{c}{{\scriptsize lang}} &
\multicolumn{1}{c}{{\scriptsize p.work}} &
 &
 &
 &
 &
 &
\\

\multicolumn{1}{c}{{\scriptsize size}} &
 &
 &
 &
 &
 &
 &
\\
\bottomrule
\end{tabular}
}

}

%-------------------------------------------------------------------------

\subsubsection{Tuning}

%      actions {
%        act = ""
%      }

We captured what kind of action is applied to a corpus during research.
In \numPapersWithModifications{} papers, sources or binaries were modified by
instrumentation, faults/clones injection, adjusting identifiers, etc.
In~\numPapersWithTestsRun{} papers, tests needed to be run against the corpus
either to verify made modifications or to collect the data.
In~\numPapersWithRun{} papers, the corpora had to be executed in order to
perform the needed analysis or to collect data.
%In~\numPapersWithBuildCompileParse{} papers, a build, or compilation, or parsing of the corpora was performed.
In~\numPapersWithContentsFiltering{} papers, some filtering of the contents of
the corpus was needed to, e.g., identify main source code/main part of
the project. 

\punchline{
We have detected few common actions applied to corpora during research:
source code/binaries modification; 
execution of the tests on the corpus or of the corpus itself;
filtering of the corpus contents.
Altogether, one out of four papers contains signs of one of these actions.

%One out of four papers with project-based corpora perform either some modification of the corpus contents or executes it directly or via tests.
%Most of its usage~(\pMaxActFromMod{}\,\%), src/bin modification gets from \confMaxActFromMod{} papers;
%most of the direct run of corpora~(\pMaxActFromRun{}\,\%) happens in \confMaxActFromRun{} papers;
%tests run~(\pMaxActFromTests{}\,\%)---in \confMaxActFromTests{} papers.

\centering
{\footnotesize
\begin{tabular}{@{}v{2em}@{\hskip 0.04in}v{2em}@{\hskip 0.04in}v{2em}@{\hskip 0.04in}v{2em}@{\hskip 0.04in}v{2em}@{\hskip 0.04in}v{2em}@{\hskip 0.04in}v{2em}@{}{r}}
\addlinespace
\addlinespace
\multicolumn{7}{c}{\textbf{Popular actions}} \\
\toprule
\multicolumn{1}{c}{ESEM} &
\multicolumn{1}{c}{MSR} &
\multicolumn{1}{c}{SCAM} &
\multicolumn{1}{c}{ICPC} &
\multicolumn{1}{c}{ICSM} &
\multicolumn{1}{c}{WCRE} &
\multicolumn{1}{c}{CSMR} &
\\

\multicolumn{1}{c}{\pMaxActInESEM{}\,\%} &
\multicolumn{1}{c}{\pMaxActInMSR{}\,\%} &
\multicolumn{1}{c}{\pMaxActInSCAM{}\,\%} &
\multicolumn{1}{c}{\pMaxActInICPC{}\,\%} &
\multicolumn{1}{r}{\pMaxActInICSM{}\,\%} &
\multicolumn{1}{c}{\pMaxActInWCRE{}\,\%} &
\multicolumn{1}{c}{\pMaxActInCSMR{}\,\%} &
\\

\multicolumn{1}{c}{{\scriptsize tests}} &
\multicolumn{1}{c}{{\scriptsize run}} &
\multicolumn{1}{c}{{\scriptsize modif.}} &
\multicolumn{1}{c}{{\scriptsize modif.}} &
\multicolumn{1}{c}{{\scriptsize tests}} &
\multicolumn{1}{c}{{\scriptsize modif.}} &
\multicolumn{1}{c}{{\scriptsize modif.}} &
\\
\bottomrule
\end{tabular}
}

}

%      manual_effort = "None"

We captured manual effort that went into creation of a
corpus, e.g., when a paper mentions setting up environments and providing needed libraries in order
to execute the corpus.
%for example, ``We made sure that we could compile and run three systems by
%setting up the appropriate environments and downloading the relevant
%libraries.''
For that, we graded each corpus on the following scale.
\emph{None}: no manual effort mentioned (\numCorporaManualNone{} corpora);
\emph{some}: some manual effort mentioned, e.g., manual detection of design patterns in source code (\numCorporaManualSomeGrouped{} corpora);
%\emph{considerable}: rather extensive effort, e.g., manual tagging of \numprint{2000} identifiers or manual classification of \numprint{6000} samples (\numCorporaManualConsiderable{} corpora);
and \emph{all} means that corpus is self-written (\numCorporaManualAll{} corpora).

\punchline{
One out of four project-based corpora requires some manual effort.
%Across conferences, this trend holds for ICPC, MSR, WCRE;
%the lowest amount of manual effort on a corpus show CSMR and ESEM papers~(\pPapersWithManualEffortCSMR{}\,\% each).

\centering
{\footnotesize
\begin{tabular}{@{}{r}@{\hskip 0.15in}{r}@{\hskip 0.15in}{r}@{\hskip 0.15in}{r}@{\hskip 0.15in}{r}@{\hskip 0.15in}{r}@{\hskip 0.15in}{r}@{}}
\addlinespace
\addlinespace
\multicolumn{7}{c}{\textbf{Manual effort}} \\
\toprule
\cPapersWithManualEffortOne{} &
\cPapersWithManualEffortTwo{} &
\cPapersWithManualEffortThree{} &
\cPapersWithManualEffortFour{} &
\cPapersWithManualEffortFive{} &
\cPapersWithManualEffortSix{} &
\cPapersWithManualEffortSeven{} 
\\

\pPapersWithManualEffortOne{}\,\% &
\pPapersWithManualEffortTwo{}\,\% &
\pPapersWithManualEffortThree{}\,\% &
\pPapersWithManualEffortFour{}\,\% &
\pPapersWithManualEffortFive{}\,\% &
\pPapersWithManualEffortSix{}\,\% &
\pPapersWithManualEffortSeven{}\,\%
\\
\bottomrule
\end{tabular}
}

}

%\pagebreak
%-------------------------------------------------------------------------
\subsection{Self-classification} 
%-------------------------------------------------------------------------

%-------------------------------------------------------------------------

\begin{wraptable}{l}{3.5cm}
{\small
\caption{Self classification}
\label{T:selfTypes}
\begin{tabular}{@{}{l}{r}@{}}
\toprule
Type & \# \\
\midrule

case study & 48 \\
experiment & 44 \\
empirical study & 22 \\
evaluation & 14 \\
exploratory study & 6 \\
... & ... \\

\bottomrule
\end{tabular}
}
\end{wraptable}

%-------------------------------------------------------------------------
 We collected explicit self-classifications
from the papers; from the sentences like ``we have conducted a case
study'' we would conclude that the current paper is a case study. Some
of the self-classifications were very detailed and precise, e.g., ``a
pre/post-test quasi experiment'', in such cases we reduced the type to
a simpler version, e.g., an experiment.  We would also count terms
like ``experimental assessment'' or ``experimental study'' towards the
experiment type.  As seen from Table~\ref{T:selfTypes}, most often
authors use terms such as ``case study'' and ``experiment'' to
describe their research.  In some cases, papers contain more than one
self-classification (\moreThanOneSelfType{} cases).  In
\noSelfTypeNum{} papers, we could not detect any self-classification.

\punchline{ Four out of five papers provide self-classification, but
  it might be vague.  The most popular term, `case study,' may be
  misused.  Cf., ``There is much confusion in the SE literature over
  what constitutes a case study.  The term is often used to mean a
  worked example. As an empirical method, a case study is something
  very different.''~\cite{EasterbrookEtAl08}.  Cf., ``... our sample
  indicated a large misuse of the term case
  study.''~\cite{ZannierEtAl06}

%Furthermore, we were able to detect twice as less experiments as was self-declared.
%Across conferences, we detected least self-classifications among SCAM papers (only~\withSelfTypeSCAMPercent{}\,\%)
%and most among ICSM papers (\withSelfTypeICSMPercent{}\,\%).

\centering
{\footnotesize
\begin{tabular}{@{}{r}@{\hskip 0.15in}{r}@{\hskip 0.15in}{r}@{\hskip 0.15in}{r}@{\hskip 0.15in}{r}@{\hskip 0.15in}{r}@{\hskip 0.15in}{r}@{}}
\addlinespace
\addlinespace
\multicolumn{7}{c}{\textbf{Self-classification}} \\
\toprule
\cPapersWithSelfTypesOne{} &
\cPapersWithSelfTypesTwo{} &
\cPapersWithSelfTypesThree{} &
\cPapersWithSelfTypesFour{} &
\cPapersWithSelfTypesFive{} &
\cPapersWithSelfTypesSix{} &
\cPapersWithSelfTypesSeven{}
\\

\pPapersWithSelfTypesOne{}\,\% &
\pPapersWithSelfTypesTwo{}\,\% &
\pPapersWithSelfTypesThree{}\,\% &
\pPapersWithSelfTypesFour{}\,\% &
\pPapersWithSelfTypesFive{}\,\% &
\pPapersWithSelfTypesSix{}\,\% &
\pPapersWithSelfTypesSeven{}\,\%
\\
\bottomrule
\end{tabular}
}

}

.
\subsection{Emerged forms}

%-------------------------------------------------------------------------

%We did not start with a pre-defined idea what a questionnaire or an experiment is.
Independently of the self-classification of the papers, we noted structural
characteristics of research performed in the papers. 
We did not use any theoretical definition for what to consider a questionnaire or an experiment.  
The developed definitions are structural, composed of the characteristics that
emerged from the papers, as they were discussed and structurally supported by
the authors.

%-------------------------------------------------------------------------

\subsubsection{Experiment}

%-------------------------------------------------------------------------

%      num_participants =

We have identified \numExperiments{} experiments in \freqExpe{}
papers.  Except for two, they all involve human subjects.  On the
average, an experiment has \medianNumExpParticipants{}
participants. The maximum number of participants is
\maxNumExpParticipants{}, the minimum is \minNumExpParticipants{},
first and third quartiles are \firstQuartileNumExpParticipants{} and
\thirdQuartileNumExpParticipants{} respectively.  In
\numExperimentsWithCorpora{} cases, an experiment uses a corpus
(in~\numExperimentsWithProjectBasedCorpora{} cases, a project-based
one); \numQsInExperiments{}~questionnaires are used in
\numExperimentsWithQ{}~experiments.

%      num_populations =
%      num_groups_participants =
%      num_tasks =
%      time_spent =

In two-thirds of the experiments, participants come from one
population, the remaining experiments draw participants from two or
three populations.  The most common source of participants is
students; sometimes distinguished by their level---graduate,
undergraduate, Bachelor, Master, and PhD students.  In one-third of
the cases, professionals are involved (full-time developers, experts,
industry practitioners, etc.). In half of the cases, participants form
the only group in the experiment. When there is more than one group
(usually, two---with a couple of exceptions of 4 and 5 groups), the
group is representing a treatment (a task), or an experience level, or
a gender.  On the average, an experiment has \medianNumExpTasks{}
tasks and lasts for an hour (with a few exceptions when an experiment
takes several weeks or even a month).

%      pilotStudy {
%        present = true
%        num_participants =
%      }
%      compensation = "None"

In \numExperimentsWithPilotStudy{} cases, it is mentioned that an
experiment had a pilot study.  In \numExperimentsWithCompensation{}
cases, it is mentioned that participants of the experiment were
offered compensation: monetary or another kind of incentive (e.g., a
box of candy).

%      requirements_explicit {
%        % Requirements for tasks
%        req = ""
%        % Requirements for participants
%        req = ""
%      }

The main requirement for the participants is their experience: basic
knowledge of used technology, or language, or IDE.  As for the tasks,
they are expected to be of a certain size (e.g., a method body to fit
on one page), or of certain contents (e.g., contain ``if''
statements).  The usual requirement for an experiment also is either
that the tested tool or used code is unfamiliar to the participants,
or on the contrary that the background is familiar (e.g., well-known
design patterns).

%
%      methodology = ""
%      methodology_quote = ""
%

\punchline{One out of ten papers contains an experiment.  The
  majority of the experiments use project-based corpora; experiments
  often use questionnaires, usually two per experiment.  An average
  experiment involves \medianNumExpParticipants{} students, often in
  two groups (by the received treatment or experience level); it
  consists of four tasks and lasts for an hour.  One out of four
  experiments suggests some compensation to its participants; one out
  of four experiments is preceded by a pilot study.

%Across conferences, experiments are rather popular in ICPC~(found~in~\pPapersWithExperimentICPC{}\,\%~papers).
ICPC and ESEM are the main source of experiments involving professionals.

\centering
{\footnotesize
\begin{tabular}{@{}{r}@{\hskip 0.15in}{r}@{\hskip 0.15in}{r}@{\hskip 0.15in}{r}@{\hskip 0.15in}{r}@{\hskip 0.15in}{r}@{\hskip 0.15in}{r}@{}}
\addlinespace
\addlinespace
\multicolumn{7}{c}{\textbf{Experiments}} \\
\toprule
\cPapersWithExperimentOne{} &
\cPapersWithExperimentTwo{} &
\cPapersWithExperimentThree{} &
\cPapersWithExperimentFour{} &
\cPapersWithExperimentFive{} &
\cPapersWithExperimentSix{} &
\cPapersWithExperimentSeven{}
\\

\pPapersWithExperimentOne{}\,\% &
\pPapersWithExperimentTwo{}\,\% &
\pPapersWithExperimentThree{}\,\% &
\pPapersWithExperimentFour{}\,\% &
\pPapersWithExperimentFive{}\,\% &
\pPapersWithExperimentSix{}\,\% &
\pPapersWithExperimentSeven{}\,\%
\\
\bottomrule
\end{tabular}
}

}

%-------------------------------------------------------------------------

\subsubsection{Questionnaire}

%-------------------------------------------------------------------------

Altogether, we have found \numQuestionnaires{} questionnaires in
\freqQues{} papers.  As mentioned, \numQsInExperiments{}
questionnaires are used in experiments---to distinguish, we will refer
to them as experiment-related and the other \numQsNotInExperiments{}
we will qualify as experiment-unrelated.

%      num_sections = 
%      num_total_questions = 
%      type_questions = % "Likert scale", "open"

Sizewise, there is no particular difference between experiment-related
and -unrelated questionnaires.  On the average, both have
\medianNumQsAll{} questions grouped in one section.
%The most popular types of questions are open questions and scale questions. 
In \numQuestionnairesWithCorpus{} cases, an experiment-unrelated
questionnaire has a corpus.

%      num_participants_asked =  
%      num_participants_answered = 
%      num_participants = 
%      num_populations =

While experiment-related questionnaires have the same participants as
the experiments they relate to (i.e., involve mostly students),
experiment-unrelated questionnaires involve professionals (testers,
managers, experts, consultants, software engineers) as participants in
two-thirds of the cases.  On the average, an experiment-unrelated
questionnaire has \medianNumQParticipants{} participants.  When it was
possible (\numQuestionnairesWithRatio{} cases), we calculated how many
participants took part in the experiment-unrelated questionnaire
compared to the initial number of questioned people. On the average,
\medianRatioQParticipants{}\,\% take part in the end, in the worst
case the ratio can be as low as~\minRatioQAllParticipants{}\,\%.

%      requirements {
While experiment-related questionnaires have the same requirements
regarding the participants as the experiments they relate to,
experiment-unrelated questionnaires have requirements concerned with
the participants' experience (e.g., Java experience) or expertise
(specific area of experience such as clone detection or web
development).

%      note = ""
When related to experiments, questionnaires are often performed before
(referred to as ``pretest'' in \numQuestionnairesPretest{}~cases) and
after the experiment (referred to as ``posttest'' in
\numQuestionnairesPosttest{}~cases).

%      pilotStudy {
In \numQuestionnairesWithPilotStudy{} cases, an experiment-unrelated questionnaire was preceded by a pilot study.

\punchline{More than half of the detected questionnaires are used in
  experiments---often as pretest and posttest questionnaires.  The
  other half, experiment-unrelated questionnaires, are found in one
  out of twelve papers.  Sizewise, on the average there is no
  difference between experiment-related and -unrelated questionnaires.
  Experiment-unrelated questionnaires usually involve professionals as
  participants---in contrast to experiment-related questionnaires that
  mostly use students.  Typical requirements for participants in
  experiment-unrelated questionnaires have to do with experience or
  expertise.  One out of three experiment-unrelated questionnaires are
  preceded by a pilot study.

%Across conferences, human surveys are rather popular in ESEM (found~in~\pMaxWexpInESEM{}\,\%~papers).
\centering
{\footnotesize
\begin{tabular}{@{}{r}@{\hskip 0.15in}{r}@{\hskip 0.15in}{r}@{\hskip 0.15in}{r}@{\hskip 0.15in}{r}@{\hskip 0.15in}{r}@{\hskip 0.15in}{r}@{}}
\addlinespace
\addlinespace
\multicolumn{7}{c}{\textbf{Experiment-unrelated questionnaires}} \\
\toprule
MSR &
CSMR &
SCAM &
WCRE &
ICSM &
ICPC &
ESEM
\\

\pMaxWexpInMSR{}\,\% &
\pMaxWexpInCSMR{}\,\% &
\pMaxWexpInSCAM{}\,\% &
\pMaxWexpInWCRE{}\,\% &
\pMaxWexpInICSM{}\,\% &
\pMaxWexpInICPC{}\,\% &
\pMaxWexpInESEM{}\,\%

\\
\bottomrule
\end{tabular}
}

}

%-------------------------------------------------------------------------

\subsubsection{Literature survey}

%-------------------------------------------------------------------------

We have found \numLitSurverys{} literature surveys in \freqLitSurv{}
papers.  Except for one, they provide extensive details on how the
survey was conducted. In particular, the used methodology is clearly
stated: four times it is said to be a ``systematic literature review''
and once a ``quasi systematic literature review''. In three cases, the
systematic literature review was done following guidelines by
Kitchenham~\cite{Kitchenham04}.

The papers are initially collected either by searching digital
libraries or from the proceedings of specific conferences and
journals.
%Conferences/journals and digital libraries are the common source of papers.
Among used digital libraries are EI Compendex, Google Scholar, ISI,
and Scopus---the latter was used in two papers. As for the conferences
and journals, there is no intersection between the lists of
names---except for ICSE, which was used in two papers.

On the average, a literature survey starts with
\medianNumLSPapersStart{} papers, its final set contains
\medianNumLSPapers{} papers, meaning that on the average only
\pSelectedPapersInLitSurverysOnAverage{}\,\% papers are taken into
account in the end.  The percentage can be as high as
\maxRatioLSPapers{}\,\% and as low as \minRatioLSPapers{}\,\%.

Requirements for papers to be included into the survey are usually
related to the scope of the investigated research. Other requirements
are concerned with the paper itself: available online, written in
English, a long paper, with empirical validation.

After all the papers are collected, they are filtered based on the
titles and abstracts, which are examined manually by the researchers
(in one case, also conclusions were taken into account; in another
case, full text of the papers was searched for keywords). Then the
full text of each paper is read and the final decision is made as to
whether to consider the paper relevant.

\punchline{ Literature surveys are quite rare: only one out of 35
  papers contains it.  On the average, a literature survey starts with
  few thousand papers to be filtered down to few dozens papers that
  will be analyzed.  Usually, the first round of filtering is based on
  the title and abstract, then the full text of the papers is
  considered.  There is not enough information to conclude about
  frequently used digital libraries or conferences/journals.  Half of
  the surveys were following guidelines of systematic literature
  reviews by Kitchenham~\cite{Kitchenham04}.

\centering
{\footnotesize
\begin{tabular}{@{}{r}@{\hskip 0.15in}{r}@{\hskip 0.15in}{r}@{\hskip 0.15in}{r}@{\hskip 0.15in}{r}@{\hskip 0.15in}{r}@{\hskip 0.15in}{r}@{}}
\addlinespace
\addlinespace
\multicolumn{7}{c}{\textbf{Literature surveys}} \\
\toprule
\cPapersWithLitSurveyOne{} &
\cPapersWithLitSurveyTwo{} &
\cPapersWithLitSurveyThree{} &
\cPapersWithLitSurveyFour{} &
\cPapersWithLitSurveyFive{} &
\cPapersWithLitSurveySix{} &
\cPapersWithLitSurveySeven{}
\\

\pPapersWithLitSurveyOne{}\,\% &
\pPapersWithLitSurveyTwo{}\,\% &
\pPapersWithLitSurveyThree{}\,\% &
\pPapersWithLitSurveyFour{}\,\% &
\pPapersWithLitSurveyFive{}\,\% &
\pPapersWithLitSurveySix{}\,\% &
\pPapersWithLitSurveySeven{}\,\%
\\
\bottomrule
\end{tabular}
}
}

%-------------------------------------------------------------------------

\subsubsection{Comparisons}

%-------------------------------------------------------------------------

During coding, we noticed the recurring motif of comparisons in the
papers.  While we did not assess the scope nor the goal, we have coded
the basic information: what is the nature of the subjects being
compared (tools, techniques), how many subjects are compared, and is
one of them introduced in the paper.

We have found comparisons in \freqComp{} papers.  Almost all of them
(except for~\numPapersWithComparisonsWOProjectBasedCorpus{}~papers),
use project-based corpora.
%Half of the time, comparison was used as means of evaluation of the suggested in the study technique, approach, or tool. 
Half of the time, a comparison is made for the technique, approach, or
tool that was introduced in the study---with the apparent reason to
evaluate the proposed technique, approach, or tool.  On the average,
such evaluation involves one other technique, approach, or tool.
In the other cases, compared were: metrics, tools, algorithms, designs,
etc.  For such comparisons, on the average, the group of compared
entities was of size \medianNumComparedWithNoCurrent{}.

\punchline{One out of three papers compares tools, techniques,
  approaches, metrics, etc.---half of the time, to evaluate what was
  introduced in the study. On the average, such evaluation involves
  one other entity.  In the other half of the cases, the average
  number of compared entities is \medianNumComparedWithNoCurrent{}.

%Comparisons are rather equally popular among all the conferences with exceptions like
%CSMR~(using comparisons~in~\pPapersWithComparisonCSMR{}\,\%~papers) and ICPC~(in~\pPapersWithComparisonICPC{}\,\%~papers).
%One out of four evaluations via comparison is done in CSMR papers.
%Two out of seven detected comparison among existing entities is done in ESEM papers. 

\centering
{\footnotesize
\begin{tabular}{@{}{r}@{\hskip 0.15in}{r}@{\hskip 0.15in}{r}@{\hskip 0.15in}{r}@{\hskip 0.15in}{r}@{\hskip 0.15in}{r}@{\hskip 0.15in}{r}@{}}
\addlinespace
\addlinespace
\multicolumn{7}{c}{\textbf{Comparisons}} \\
\toprule
\cPapersWithComparisonOne{} &
\cPapersWithComparisonTwo{} &
\cPapersWithComparisonThree{} &
\cPapersWithComparisonFour{} &
\cPapersWithComparisonFive{} &
\cPapersWithComparisonSix{} &
\cPapersWithComparisonSeven{}
\\

\pPapersWithComparisonOne{}\,\% &
\pPapersWithComparisonTwo{}\,\% &
\pPapersWithComparisonThree{}\,\% &
\pPapersWithComparisonFour{}\,\% &
\pPapersWithComparisonFive{}\,\% &
\pPapersWithComparisonSix{}\,\% &
\pPapersWithComparisonSeven{}\,\%
\\
\bottomrule
\end{tabular}
}
}

%-------------------------------------------------------------------------
\begin{table}
\begin{threeparttable}
\caption{Existing tools used in the papers}
\label{T:tools}

\begin{tabular}{lr}

\begin{minipage}{.45\linewidth}
\centering

\begin{tabular}{@{}v{7em}@{ }{r}@{}}
\toprule
Tool & \# papers \\
\midrule

Eclipse\tnote{1} & \numPapersToolEclipse{} \\
R project\tnote{2} & \numPapersToolR{} \\
CCFinder\tnote{3} & \numPapersToolCCFinder{} \\
Understand\tnote{4} & \numPapersToolUnderstand{} \\
Weka\tnote{5} & \numPapersToolWeka{} \\
ConQAT\tnote{6} & \numPapersToolConqat{} \\

\bottomrule
\end{tabular}

\end{minipage}%

&

\begin{minipage}{.45\linewidth}
\centering

\begin{tabular}{@{}v{7em}@{ }{r}@{}}
\toprule
Tool & \# papers \\
\midrule

MALLET\tnote{7} & \numPapersToolMallet{} \\
ChangeDistiller\tnote{8} & \numPapersToolChangedistiller{} \\
CodeSurfer\tnote{9} & \numPapersToolCodesurfer{} \\
Evolizer\tnote{10} & \numPapersToolEvolizer{} \\
RapidMiner\tnote{11} & \numPapersToolRapidminer{} \\
RECODER\tnote{12} & \numPapersToolRecoder{} \\

\bottomrule
\end{tabular}

\end{minipage}%

\end{tabular}
\begin{tablenotes}
\item [1] \url{http://eclipse.org}
\item [2] \url{http://www.r-project.org/}
\item [3] \url{http://www.ccfinder.net/}
\item [4] \url{http://www.scitools.com/}
\item [5] \url{http://www.cs.waikato.ac.nz/ml/weka/}
\item [6] \url{https://www.conqat.org/}
\item [7] \url{http://mallet.cs.umass.edu/}
\item [8] \url{http://www.ifi.uzh.ch/seal/research/tools/changeDistiller.htm}
\item [9] \url{http://www.grammatech.com/products/codesurfer/overview.html}
\item [10] \url{http://www.ifi.uzh.ch/seal/research/tools/evolizer.html}
\item [11] \url{http://rapid-i.com/content/view/181/190/}
\item [12] \url{http://sourceforge.net/projects/recoder/}
\end{tablenotes}
\end{threeparttable}

\end{table}

%-------------------------------------------------------------------------
%-------------------------------------------------------------------------
\subsection{Tools}
%-------------------------------------------------------------------------

We have found \numPapersWithNameTools{} papers to introduce a tool
(where we were able to capture this fact only if the name of the tool
was mentioned or it was clearly stated that ``a prototype'' is
implemented).  In \numPapersWithNonameTools{} more papers, we detected
that additional, helper tooling for the current purpose of the study
is implemented (parsers, analyzers, and so on).

When names of existing tools were explicitly mentioned to be used, we
collected the names.  We have found that in
\numPapersWithToolsNoIntroducedNoNoname{} cases, a paper makes use of
existing tools.  On the average, a paper uses \medianNumToolsInPaper{}
tools; the captured maximum is \maxNumToolsInPaper{}.  The frequently
used tools are listed in~Table~\ref{T:tools}.  We counted towards
Eclipse usage also cases when a paper used an existing tool that we
know to be an Eclipse plug-in.  For brevity, we omit names of
\numToolsUsedInTwoPapers{} tools each of which was used in two papers.

\punchline{ One out of four papers introduces a new tool; another one
  out of four papers uses some home-grown tooling.  Almost three out
  of four papers use existing tools.

  The most popular standard tool, Eclipse---an IDE and a platform for
  plug-in development---is used in one out of seven papers.  Other
  popular tools cater for source code analysis, clone detection,
  evolution analysis, data mining, statistics, quality analysis,
  document classification.

\centering
{\footnotesize
\begin{tabular}{@{}{r}@{\hskip 0.15in}{r}@{\hskip 0.15in}{r}@{\hskip 0.15in}{r}@{\hskip 0.15in}{r}@{\hskip 0.15in}{r}@{\hskip 0.15in}{r}@{}}
\addlinespace
\addlinespace
\multicolumn{7}{c}{\textbf{Home-grown tooling}} \\
\toprule
\cPapersWithNonameToolsOne{} &
\cPapersWithNonameToolsTwo{} &
\cPapersWithNonameToolsThree{} &
\cPapersWithNonameToolsFour{} &
\cPapersWithNonameToolsFive{} &
\cPapersWithNonameToolsSix{} &
\cPapersWithNonameToolsSeven{}
\\

\pPapersWithNonameToolsOne{}\,\% &
\pPapersWithNonameToolsTwo{}\,\% &
\pPapersWithNonameToolsThree{}\,\% &
\pPapersWithNonameToolsFour{}\,\% &
\pPapersWithNonameToolsFive{}\,\% &
\pPapersWithNonameToolsSix{}\,\% &
\pPapersWithNonameToolsSeven{}\,\%
\\
\bottomrule
\end{tabular}
}

\centering
{\footnotesize
\begin{tabular}{@{}{r}@{\hskip 0.15in}{r}@{\hskip 0.15in}{r}@{\hskip 0.15in}{r}@{\hskip 0.15in}{r}@{\hskip 0.15in}{r}@{\hskip 0.15in}{r}@{}}
\addlinespace
\addlinespace
\multicolumn{7}{c}{\textbf{Introduced tools}} \\
\toprule
\cPapersWithIntroducedToolsOne{} &
\cPapersWithIntroducedToolsTwo{} &
\cPapersWithIntroducedToolsThree{} &
\cPapersWithIntroducedToolsFour{} &
\cPapersWithIntroducedToolsFive{} &
\cPapersWithIntroducedToolsSix{} &
\cPapersWithIntroducedToolsSeven{}
\\

\pPapersWithIntroducedToolsOne{}\,\% &
\pPapersWithIntroducedToolsTwo{}\,\% &
\pPapersWithIntroducedToolsThree{}\,\% &
\pPapersWithIntroducedToolsFour{}\,\% &
\pPapersWithIntroducedToolsFive{}\,\% &
\pPapersWithIntroducedToolsSix{}\,\% &
\pPapersWithIntroducedToolsSeven{}\,\%
\\
\bottomrule
\end{tabular}
}

}

%-------------------------------------------------------------------------

\subsection{Structural signs of rigorousness/quality}

We do not aim to assess the quality or rigorousness of the studies.
We capture presence of some of the aspects that are taken into account
when assessing rigorousness/quality of research
(cf.,~\cite{IvarssonG11})---in that, we restrict ourselves only to the
structural aspects.

%-------------------------------------------------------------------------
\subsubsection{Study presentation aspects}
%-------------------------------------------------------------------------

A clear set of definitions for the terms used in the paper is found in
\numPapersWithDefinitions{} papers.  Research questions are adopted in
\numPapersWithResearchQuestions{} papers.  In \numPapersWithGQM{}
papers, a ``Goal-Question-Metric'' approach is used.  Explicit
mention of null hypothesis or hypotheses is found in
\numPapersWithHypotheses{} papers.  Section ``Threats to validity'' is
present in \numPapersWithThreatsSection{} papers; of them,
\numPapersWithWohlinThreats{} discuss threats using classification
described, e.g., in \cite{WohlinEtAl00}: threats to external
(mentioned in \numPapersWithExternal{} papers), internal
(\numPapersWithInternal{} papers), construct
(\numPapersWithConstruct{} papers), and conclusion
(\numPapersWithConclusion{} papers) validity.

If to consider combinations of these signs (definitions, research
questions, hypotheses, and threats), the most popular one is the
absence of all of them: demonstrated
by~\numPapersWithoutAnything{}~papers.  The second most popular
combination is presence of research questions and threats to validity:
found in~\numPapersWithRQsAndThreats{}~papers.  The third most
popular---usage of only threats to validity---found
in~\numPapersWithOnlyThreats{}~papers.  Together, these three
combinations
describe~\pPapersThreeTopCombinations{}\,\%~of~the~papers.

\punchline{ Half of the papers use research questions to structure
  their study.
%Across conferences, this style is highly adopted in ESEM~papers~(\pPapersWithRQESEM{}\,\%), the lowest adoption is in SCAM~papers~(\pPapersWithRQSCAM{}\,\%). 
  One out of seven papers uses a ``Goal-Question-Metric'' approach and/or
  formulate (null) hypotheses to structure their research. One out of
  seven papers provides an explicit set of definitions of the terms
  used in the study.
%Across conferences, ``Goal-Question-Metric'' is adopted the most in ICPC~papers~(\pPapersWithGQMICPC{}\,\%), hypotheses are often found in MSR~papers~(\pPapersWithHypothesesMSR{}\,\%), and definitions---in~WCRE~papers~(\pPapersWithDefinitionsWCRE{}\,\%).
  Threats to validity are discussed in three out of five papers.
%This is also the trend across conferences with exceptionally high~(\pPapersWithThreatsICPC{}\,\%)
%and low~(\pPapersWithThreatsSCAM{}\,\%) percentage of papers in ICPC and SCAM respectively.

%Across conferences, the most popular combination---absence of all the structural signs---is noticed the most in SCAM papers~(\pPapersWithNoQualitySignsSCAM{}\,\%)
%and the least in ICPC papers~(\pPapersWithNoQualitySignsICPC{}\,\%).
%The second most popular combination---research questions and threats to
%validity---is rather adopted among ESEM
%papers~(\pPapersWithRQsAndThreatsESEM{}\,\%). The third most popular
%combination---threats to validity only---is rather common among ICSM
%papers~(\pPapersWithOnlyThreatsICSM{}\,\%).

%For all conferences, these three most popular combinations cover at least half of the papers, except for WCRE, where only~\pPapersWithThreeTopQualityTypesWCRE{}\,\%~of~papers are described by these combinations.

  The following three combinations of structural signs describe at
  least half of the papers in each conference, except for WCRE, where
  only~\pPapersWithThreeTopQualityTypesWCRE{}\,\%~of~papers are
  covered by these combinations.

\centering
{\footnotesize
\begin{tabular}{@{}{r}@{\hskip 0.15in}{r}@{\hskip 0.15in}{r}@{\hskip 0.15in}{r}@{\hskip 0.15in}{r}@{\hskip 0.15in}{r}@{\hskip 0.15in}{r}@{}}
\addlinespace
\addlinespace
\multicolumn{7}{c}{\textbf{No structural signs}} \\
\toprule
\cPapersWithNoQualitySignsOne{} &
\cPapersWithNoQualitySignsTwo{} &
\cPapersWithNoQualitySignsThree{} &
\cPapersWithNoQualitySignsFour{} &
\cPapersWithNoQualitySignsFive{} &
\cPapersWithNoQualitySignsSix{} &
\cPapersWithNoQualitySignsSeven{}
\\

\pPapersWithNoQualitySignsOne{}\,\% &
\pPapersWithNoQualitySignsTwo{}\,\% &
\pPapersWithNoQualitySignsThree{}\,\% &
\pPapersWithNoQualitySignsFour{}\,\% &
\pPapersWithNoQualitySignsFive{}\,\% &
\pPapersWithNoQualitySignsSix{}\,\% &
\pPapersWithNoQualitySignsSeven{}\,\%
\\
\bottomrule
\end{tabular}
}

\centering
{\footnotesize
\begin{tabular}{@{}{r}@{\hskip 0.15in}{r}@{\hskip 0.15in}{r}@{\hskip 0.15in}{r}@{\hskip 0.15in}{r}@{\hskip 0.15in}{r}@{\hskip 0.15in}{r}@{}}
\addlinespace
\addlinespace
\multicolumn{7}{c}{\textbf{Both research questions and threats to validity}} \\
\toprule
\cPapersWithRQsAndThreatsOne{} &
\cPapersWithRQsAndThreatsTwo{} &
\cPapersWithRQsAndThreatsThree{} &
\cPapersWithRQsAndThreatsFour{} &
\cPapersWithRQsAndThreatsFive{} &
\cPapersWithRQsAndThreatsSix{} &
\cPapersWithRQsAndThreatsSeven{}
\\

\pPapersWithRQsAndThreatsOne{}\,\% &
\pPapersWithRQsAndThreatsTwo{}\,\% &
\pPapersWithRQsAndThreatsThree{}\,\% &
\pPapersWithRQsAndThreatsFour{}\,\% &
\pPapersWithRQsAndThreatsFive{}\,\% &
\pPapersWithRQsAndThreatsSix{}\,\% &
\pPapersWithRQsAndThreatsSeven{}\,\%
\\
\bottomrule
\end{tabular}
}

\centering
{\footnotesize
\begin{tabular}{@{}{r}@{\hskip 0.15in}{r}@{\hskip 0.15in}{r}@{\hskip 0.15in}{r}@{\hskip 0.15in}{r}@{\hskip 0.15in}{r}@{\hskip 0.15in}{r}@{}}
\addlinespace
\addlinespace
\multicolumn{7}{c}{\textbf{Only threats to validity}} \\
\toprule
\cPapersWithOnlyThreatsOne{} &
\cPapersWithOnlyThreatsTwo{} &
\cPapersWithOnlyThreatsThree{} &
\cPapersWithOnlyThreatsFour{} &
\cPapersWithOnlyThreatsFive{} &
\cPapersWithOnlyThreatsSix{} &
\cPapersWithOnlyThreatsSeven{}
\\

\pPapersWithOnlyThreatsOne{}\,\% &
\pPapersWithOnlyThreatsTwo{}\,\% &
\pPapersWithOnlyThreatsThree{}\,\% &
\pPapersWithOnlyThreatsFour{}\,\% &
\pPapersWithOnlyThreatsFive{}\,\% &
\pPapersWithOnlyThreatsSix{}\,\% &
\pPapersWithOnlyThreatsSeven{}\,\%
\\
\bottomrule
\end{tabular}
}

}

%-------------------------------------------------------------------------
\subsubsection{Validation}
%-------------------------------------------------------------------------

We captured the mentions of performed validation of done research.  We
have found evidence of some kind of validation in
\numPapersWithSomeValidation{} papers.  In
\numPapersWithManualValidation{} cases, validation was manually
performed: either the results are small enough, or a sufficient subset
is checked.  In \numPapersWithValidationAgainst{} cases, validation
was done against existing or prepared results: actual data (when
evaluating predictions), data from previous work, or an oracle/gold
standard.  In \numPapersWithCrossValidation{} cases, cross-validation
was used.

%-------------------------------------------------------------------------

\subsection{Reproducibility}

%-------------------------------------------------------------------------

We looked for signs of additionally provided data for a replication of
the study.  Since it is usually done via the Internet, we searched the
papers for (the stems of) the following keywords: ``available,''
``download,'' ``upload,'' ``reproduce,'' ``replicate,'' ``host,'',
``URL,'' ``website,'' ``http,'' ``html''.  In such manner, we have
found links in \numPapersWithProvidedReplicationPackage{} papers.  In
\numReplicationLinksLeadingToNothing{} cases, we could not find any
mentioned material, tools or data,---links led to a general page or to
a homepage, which we searched thoroughly but without success.  In
\numReplicationLinksWrong{} more cases, we have found replication
material on the website after some searching.

\punchline{One out of three papers additionally provides online some
  data from the study, though not always to be found.

%Across conferences, the practice of uploading online some data from the study is rather equally adopted, 
%with the highest adoption in ICPC papers (found in~\pPapersWithRepPackageICPC{}\,\%~papers) and lowest---in SCAM~(\pPapersWithRepPackageSCAM{}\,\%~papers).

\centering
{\footnotesize
\begin{tabular}{@{}{r}@{\hskip 0.15in}{r}@{\hskip 0.15in}{r}@{\hskip 0.15in}{r}@{\hskip 0.15in}{r}@{\hskip 0.15in}{r}@{\hskip 0.15in}{r}@{}}
\addlinespace
\addlinespace
\multicolumn{7}{c}{\textbf{Additional data provided}} \\
\toprule
\cPapersWithRepPackageOne{} &
\cPapersWithRepPackageTwo{} &
\cPapersWithRepPackageThree{} &
\cPapersWithRepPackageFour{} &
\cPapersWithRepPackageFive{} &
\cPapersWithRepPackageSix{} &
\cPapersWithRepPackageSeven{}
\\

\pPapersWithRepPackageOne{}\,\% &
\pPapersWithRepPackageTwo{}\,\% &
\pPapersWithRepPackageThree{}\,\% &
\pPapersWithRepPackageFour{}\,\% &
\pPapersWithRepPackageFive{}\,\% &
\pPapersWithRepPackageSix{}\,\% &
\pPapersWithRepPackageSeven{}\,\%
\\
\bottomrule
\end{tabular}
}

}

As to the nature of the provided data, in \numProvidedTools{} cases,
an introduced tool or tooling used in the research is provided.  In
\numProvidedCorpus{} cases, the used corpus---in full or
partially---is provided; the complete description of the corpus (list
of used projects with their versions and/or links) is provided by
\numProvidedDescOfCorpus{} papers.  Raw data is available for
\numProvidedRawData{} papers; the same number of papers provide final
or/and additional results of the study.

% versions

When the corpus is not provided by the paper, but the names of the
used projects are mentioned, the main aspect of being able to
reproduce the corpus is knowing which versions of the projects were
used.  We noticed that in \numPapersWithNoVersions{} papers versions
of the used projects are not provided.  In \numPapersWithVersions{}
papers, versions of the projects are mentioned explicitly; in
\numPapersWithTimePeriods{} more cases, it is possible to reconstruct
the version from the mentioned time periods that the study spans.

% can_reproduce

%As Brooks et al. state in the chapter on replication~\cite{BrooksEtAl08} in the
%textbook on advanced empirical software engineering, ``Exact replication is unattainable.'' 

Altogether, we judged \numPapersReproducible{} papers to be
reproducible, meaning that either all components were provided by the
authors or we concluded that the paper contains enough details to
collect exactly the same corpus and the same tools. We did \emph{not}
judge if it is possible to follow the provided instructions, specific
to the reported research.

\begin{comment}
Often, 
In other cases,  can be done modulo manual effort or modulo missing versions of the projects, or modulo closed source. 
In many cases, it is also a question of the implementation, whether the used
home-tooling is available or re-implementable from the description given in the paper;
the same stands for introduced approaches, techniques and tools---whether they
are available (e.g., as prototypes) or re-implementable. 
\end{comment}

We also would like to note that \numPapersWithReplication{} papers mention that
they are doing a replication in their study, of them
\numPapersWithSelfReplication{} papers with self-replication.

\punchline{We judged one out of six papers to be reproducible with
  respect to the used corpus and tools.  We did \emph{not} assess
  whether enough details were provided to re-conduct the research
  itself.

\centering
{\footnotesize
\begin{tabular}{@{}{r}@{\hskip 0.15in}{r}@{\hskip 0.15in}{r}@{\hskip 0.15in}{r}@{\hskip 0.15in}{r}@{\hskip 0.15in}{r}@{\hskip 0.15in}{r}@{}}
\addlinespace
\addlinespace
\multicolumn{7}{c}{\textbf{Judged to be reproducible}} \\
\toprule
\cPapersCanReproduceOne{} &
\cPapersCanReproduceTwo{} &
\cPapersCanReproduceThree{} &
\cPapersCanReproduceFour{} &
\cPapersCanReproduceFive{} &
\cPapersCanReproduceSix{} &
\cPapersCanReproduceSeven{}
\\

\pPapersCanReproduceOne{}\,\% &
\pPapersCanReproduceTwo{}\,\% &
\pPapersCanReproduceThree{}\,\% &
\pPapersCanReproduceFour{}\,\% &
\pPapersCanReproduceFive{}\,\% &
\pPapersCanReproduceSix{}\,\% &
\pPapersCanReproduceSeven{}\,\%
\\
\bottomrule
\end{tabular}
}

%The highest percentage of such papers is found in MSR~(\pPapersCanReproduceMSR{}\,\%~papers), 
%the lowest---in WCRE~(\pPapersCanReproduceWCRE{}\,\%~papers) and ICSM~(\pPapersCanReproduceICSM{}\,\%).

%Papers doing a replication in their study are rather rare. We have found them
%only in CSMR, ESEM, ICSM, and WCRE---in each conference, the percentage of
%replications is rather low:
%from~\pPapersReplicationsThemselvesWCRE{}\,\%~to~\pPapersReplicationsThemselvesESEM{}\,\%
}

%-------------------------------------------------------------------------

\subsection{Assessment}

%-------------------------------------------------------------------------

% table, confidence

Though usually information we extracted from the papers was scattered across different sections, 
half of the papers had tables (listing projects, their names, versions, used
releases, and similar information) that helped us during coding.
We captured our confidence in the coded profile of each paper. For that, we used the following scale:
high, moderate, and low level of confidence. 
The results are as follows: 
%high---\numPapersReasHigh{} papers, 
%reasonably high---\numPapersHighEnough{} papers, 
high---\numPapersHighGrouped{} papers,
moderate---\numPapersModerate{} papers, 
%rather low---\numPapersRathLow{} papers, 
%low---\numPapersLow{} papers.
low---\numPapersLowGrouped{} papers.

\punchline{We have low confidence in one out of eleven papers that we
  have coded.  In the rest, half of the time we are moderately
  confident and half of the time---highly confident in the results.

%Across conferences, we are mostly highly confident in coding of
%CSMR~(\pMaxConfidenceInCSMR{}\,\%) and ICPC~(\pMaxConfidenceInICPC{}\,\%)
%papers.
%We have reasonably high confidence in coding of MSR papers~(\pMaxConfidenceInMSR{}\,\%);
%coding of ESEM papers has equally frequently~(\pMaxConfidenceInESEM{}\,\%) our moderate and high confidence;
%finally, we are moderately confident in coding of ICSM~(\pMaxConfidenceInICSM{}\,\%), 
%SCAM~(\pMaxConfidenceInSCAM{}\,\%), 
%and WCRE~(\pMaxConfidenceInWCRE{}\,\%) papers. 

\centering
{\footnotesize
\begin{tabular}{@{}{r}@{\hskip 0.15in}{r}@{\hskip 0.15in}{r}@{\hskip 0.15in}{r}@{\hskip 0.15in}{r}@{\hskip 0.15in}{r}@{\hskip 0.15in}{r}@{}}
\addlinespace
\addlinespace
\multicolumn{7}{c}{\textbf{High confidence}} \\
\toprule
\cPapersWithHighConfidenceLevelOne{} &
\cPapersWithHighConfidenceLevelTwo{} &
\cPapersWithHighConfidenceLevelThree{} &
\cPapersWithHighConfidenceLevelFour{} &
\cPapersWithHighConfidenceLevelFive{} &
\cPapersWithHighConfidenceLevelSix{} &
\cPapersWithHighConfidenceLevelSeven{}
\\

\pPapersWithHighConfidenceLevelOne{}\,\% &
\pPapersWithHighConfidenceLevelTwo{}\,\% &
\pPapersWithHighConfidenceLevelThree{}\,\% &
\pPapersWithHighConfidenceLevelFour{}\,\% &
\pPapersWithHighConfidenceLevelFive{}\,\% &
\pPapersWithHighConfidenceLevelSix{}\,\% &
\pPapersWithHighConfidenceLevelSeven{}\,\%
\\
\bottomrule
\end{tabular}
}

\centering
{\footnotesize
\begin{tabular}{@{}{r}@{\hskip 0.15in}{r}@{\hskip 0.15in}{r}@{\hskip 0.15in}{r}@{\hskip 0.15in}{r}@{\hskip 0.15in}{r}@{\hskip 0.15in}{r}@{}}
\addlinespace
\addlinespace
\multicolumn{7}{c}{\textbf{Moderate confidence}} \\
\toprule
\cPapersWithModerateConfidenceLevelOne{} &
\cPapersWithModerateConfidenceLevelTwo{} &
\cPapersWithModerateConfidenceLevelThree{} &
\cPapersWithModerateConfidenceLevelFour{} &
\cPapersWithModerateConfidenceLevelFive{} &
\cPapersWithModerateConfidenceLevelSix{} &
\cPapersWithModerateConfidenceLevelSeven{}
\\

\pPapersWithModerateConfidenceLevelOne{}\,\% &
\pPapersWithModerateConfidenceLevelTwo{}\,\% &
\pPapersWithModerateConfidenceLevelThree{}\,\% &
\pPapersWithModerateConfidenceLevelFour{}\,\% &
\pPapersWithModerateConfidenceLevelFive{}\,\% &
\pPapersWithModerateConfidenceLevelSix{}\,\% &
\pPapersWithModerateConfidenceLevelSeven{}\,\%
\\
\bottomrule
\end{tabular}
}
\centering
{\footnotesize
\begin{tabular}{@{}{r}@{\hskip 0.15in}{r}@{\hskip 0.15in}{r}@{\hskip 0.15in}{r}@{\hskip 0.15in}{r}@{\hskip 0.15in}{r}@{\hskip 0.15in}{r}@{}}
\addlinespace
\addlinespace
\multicolumn{7}{c}{\textbf{Low confidence}} \\
\toprule
\cPapersWithLowConfidenceLevelOne{} &
\cPapersWithLowConfidenceLevelTwo{} &
\cPapersWithLowConfidenceLevelThree{} &
\cPapersWithLowConfidenceLevelFour{} &
\cPapersWithLowConfidenceLevelFive{} &
\cPapersWithLowConfidenceLevelSix{} &
\cPapersWithLowConfidenceLevelSeven{}
\\

\pPapersWithLowConfidenceLevelOne{}\,\% &
\pPapersWithLowConfidenceLevelTwo{}\,\% &
\pPapersWithLowConfidenceLevelThree{}\,\% &
\pPapersWithLowConfidenceLevelFour{}\,\% &
\pPapersWithLowConfidenceLevelFive{}\,\% &
\pPapersWithLowConfidenceLevelSix{}\,\% &
\pPapersWithLowConfidenceLevelSeven{}\,\%
\\
\bottomrule
\end{tabular}
}

}

%-------------------------------------------------------------------------

\section{Related work}
\label{S:rel}
\setcounter{subsubsection}{0}

We summarized related work---in the sense of other literature surveys
on SE research---in the introduction and Table~\ref{T:rel}. Thus, the
key differences between our survey and previous work are these: i)
predefined schema in previous work versus emerged schema in the
present survey; ii) focus on specific forms or characteristics of SE
research in previous work versus broad analysis of empirical evidence
in the present survey. Below we compare the findings of related work
where they overlap with ours.

As a general remark, we believe that our findings quantitatively
differ from previous findings because of several factors: i) the
dependence on the choice of venues: even conferences in our study
differ considerably; ii) passed time: there is at least a five-year gap,
during which popularity of empirical research and of its
particular forms might have grown; iii) the cited papers use mostly
journals: this may increase the aforementioned gap because of the
longer process for journal publications; iv) snapshot versus
longitudinal approach: we take into account all papers of the latest
proceedings while the cited papers focus on a sample across several
years.

%-------------------------------------------------------------------------

The closest work to ours is by Zannier et al.~\cite{ZannierEtAl06}: they measured quantity
and quality of empirical evaluation in ICSE papers over the years. Our
work provides a snapshot study aiming to represent SE research broadly
across conferences. Zannier et al.\
%have not found any empirical evaluation in 19 papers of the sample; 
when assigning types to the papers, could confirm the
self-classification of half of the studies. Which agrees with our
observation that self-classification is rather weak among SE papers.
%If to understand under `empirical evaluation'
They also observe the extremely low usage of hypotheses (only one
paper) and absence of replications.  We do find some adoption of null
hypotheses and replications.
%We attribute it to the increased quality of methodology of SE research.
 
According to their classification, Glass et al.~\cite{GlassVR02} have
found~1.1\,\%~papers to contain literature reviews and~3\,\%~papers to
present ``laboratory experiment (human subjects).''  We also discover
that number of literature surveys and experiments is low, but
relatively it increased 2-3 times.
%It is interesting to note that in four out of five applied classification schemes, 
%Glass et al. found that one category describes from 49\,\% to 98\,\% of surveyed papers. 

Kitchenham et al.~\cite{KitchenhamEtAl09} considered only 0.75\% of surveyed papers to be systematic literature reviews. 
%~19 papers, i.e.,
We have found literature surveys
in~\freqLitSurv{}~papers, one of which did not contain a clear
methodology---a requirement to be met by Kitchenham's inclusion
criteria---leaving 4 papers.  Thus, our percentage of detected
literature surveys is~2.3\,\%
%---three times higher. 
\begin{comment}
  We see the explanation in two facts. First, the original guidelines
  for systematic reviews appeared only in 2004~\cite{Kitchenham04}
  (i.e., Kitchenham discovered low adoption because only three years
  passed). Second, three papers that we discovered come from ESEM ---a
  venue probably most attractive to such studies (though, in 2004-2007
  years, Kitchenham found only 1 relevant paper in ESEM).
  Kitchenham's study being a literature survey on its own falls within
  our observed range of ratio for included papers.
\end{comment}

%\kate{Make it all either present or past tense: ``detect'' or ``detected''}

%Sj{\o}berg et al. detect 113 controlled experiments in 103
% papers---which is approximately the same ratio `experiment2paper' as
% discovered by us.

According to Sj{\o}berg et al.'s study~\cite{SjobergEtAl05}, only~2{}\,\% of the papers
contain experiments, while we discover~10\,\% surveyed papers to
contain an experiment.
%Again, we mainly attribute this difference to the time gap. We also did not
%make a difference between controlled and quasi-experiments. 
On the average, Sj{\o}berg et al.\ detected an experiment to involve 30
participants---in~72.6\,\%~cases only students, in~18.6\,\%~cases only
professionals, and~in~8\,\%~cases mixed groups.  We have found that on
the average an experiment involves 16 participants---in~57\,\%~cases
only students, in~14\,\%~cases only professionals, and~in~29\,\%~cases
mixed groups.

%-------------------------------------------------------------------------

%-------------------------------------------------------------------------

\section{Threats to validity}
\label{S:threats}

%-------------------------------------------------------------------------

\setcounter{subsubsection}{0}

\subsubsection{Choice of the papers}

We did not use journal articles---while they might provide more
information or be of higher quality, we wanted to capture the state of
the common research, of which we believe conference proceedings to be
more representative.  We have chosen conferences with proceedings of
similar and reasonable size: so that not to skew the general results
by one larger conference and so that to include all the papers but
still be able to process them within reasonable period of time.
% (e.g., we managed to process from 7 to 10 papers a day).
Specifically, we excluded the ICSE conference, which had 87 long
papers in the proceedings of 2012 edition. Altogether, this means our
results might not be generalizable, but we believe them to be
representative enough.

\subsubsection{Choice of the period} 

Since we perform a snapshot study, it might be that some of the
discovered numbers are a coincidental spike.  A longitudinal
study---possible future work---may provide more details and deeper
understanding.

\subsubsection{Coding} 

The effort was manual with occasional search by specific keywords
(mentioned in the appropriate subsections of Section~\ref{S:res}).  In
\numPapersNonSearchablePDFs{} cases, papers were OCR-scanned.
% (i.e., non-searchable).

\textbf{Human factor.} Coding was done by one researcher, but the
results of the first pass were cross-validated during the second pass
as well as during the aggregation phase.  When in doubt, the
researcher constantly referred back to the surveyed papers to
double-check.

\textbf{Scheme.}  We do not claim our coding scheme to be complete or
advanced. We captured basic data related to the used empirical
evidence, often either obvious or structurally supported.  Therefore,
we might miss sophisticated or under-specified forms of empirical
research. % E.g., mostly, we detected experiments with human subjects.

%-------------------------------------------------------------------------

%-------------------------------------------------------------------------

\section{Conclusion}
\label{S:concl}

\setcounter{subsubsection}{0}

In this paper, we presented a literature survey on empirical evidence
in Software Engineering research.

\subsubsection*{Answers to the research questions}

Coming back to the initial questions that motivated our
research~(see~Section~\ref{S:meth}), we suggest the following answers:
\begin{enumerate}[I]

%\item How often SE papers use \emph{corpora}---collections of empirical evidence?
\item The overwhelming majority of surveyed conference papers use
  corpora---collections of empirical evidence.

%\item What is the nature and characteristics of the used corpora?
\item The majority of the corpora consist of projects and can be
  characterized by size, code form, software language, evolution
  measures, requirements, and applied tunings.

%\item Is it possible to identify common contents between the used corpora?
\item There are no frequently used projects or corpora across all the
  papers. We have detected though some pattern of project recurrence
  with low frequency.
  % This is the best answer that we can give based on the material of
  % the current study.
\end{enumerate}
In what follows, we further interpret these findings.

\subsubsection*{No ``holy grail''} 

Though corpora are used in the majority of the surveyed papers and
some clusters of characteristics of the used corpora are recurrent
(e.g., the use of many open source Java projects), the usage of
established datasets is low. We suggest two possible reasons. First,
adoption may be low only yet: among detected datasets being
used~(see~Table~\ref{T:setsrepos}), the oldest dataset, SIR, was
introduced in 2005, the youngest, Qualitas---in~2010.  Second,
researchers may prefer to collect and prepare their corpora
themselves, because there might not be a ``holy grail'' among corpora
to suit all possible needs. Partially, this assumption is supported
by the fact that, even on the level of projects, no clear favorite was
detected among the papers. The emerged schema with its components for
requirements and tunings for corpora also substantiates indeed the
different needs of research efforts.

\subsubsection*{Community-specific curated collections} 

On the other hand, we find that three out of seven conferences have
favorite projects, when considered separately---projects that are used
by a quarter of the papers within these conferences.
%The possible reason for that is most probably the fact that authors tend to
%re-use systems previously introduced in their field of research---e.g., for
%possible comparison of results. 
This leads to a refined version of the third question in our study:
When it is possible to detect commonly used projects within a
conference, would it be useful to provide a curated version of them?
Generally, it is clear that even requirements and tunings are
recurrent across research efforts, and hence, some ``product line'' of
curated collections and some discipline of ``corpus engineering'' may
ultimately lead to more reuse of empirical evidence. These are topics
for future work.

\begin{comment}
\subsubsection{Structure of the papers and secondary studies}
We have found that literature surveys and replications are rather rare.
We assume that one reason for that might be the amount of work to extract needed data from the papers---in the end always manual work. 
It is tedious and laborious task, needed to be done with caution and care. 
From our experience, it helps greatly, when the information that the researcher
looks for is structured in an obvious-to-eye manner: tables, bolded or emphasized words,
separate sections. 
Therefore, we believe that recognizing existing forms of SE research and
imposing specific structural requirements can help to promote secondary empirical
research as well as to increase quality of primary empirical research.
\end{comment}

\subsubsection*{Top-down vs. bottom-up introduction of methodology}

While there is a need for adoption of advanced and theoretically
specified forms of empirical research, we believe that there is a
certain amount of de facto empirical research in Software Engineering
that has formed historically. This survey sought to understand the
characteristics of empirical evidence in research---also to enable
assessment, if not improvement, of research quality. Future work
includes aligning research areas or goals with the kind of used
empirical evidence: deeper understanding of the needs may provide
insights for streamlining research.

%-------------------------------------------------------------------------

\begin{comment}
\section*{Acknowledgment} 
We would like to thank all the authors who kindly
shared with us their copies of publications when we could not access them
otherwise.
\end{comment}

%-------------------------------------------------------------------------

\bibliographystyle{IEEEtran}
\bibliography{IEEEabrv,paper}

%-------------------------------------------------------------------------

\end{document}